%% file: onestep_arxiv_ec.tex
\documentclass[12pt]{article}
\usepackage{longtable}
\usepackage[section]{placeins}
\usepackage[top=1in, left=1in, right=1in, bottom=1in]{geometry}
\usepackage[parfill]{parskip}	
\usepackage{graphicx}
\usepackage{caption}
\usepackage{ulem}
\usepackage{subcaption} 
\usepackage{mwe} 
\usepackage{cancel} 
\usepackage{soul} 
\usepackage[export]{adjustbox}
\usepackage{amssymb}
\usepackage{epstopdf}
\usepackage{bbm}
\usepackage{bm}
\usepackage{amssymb}
\usepackage{amsmath}
\usepackage{xcolor}
\usepackage{amsfonts}

\usepackage{lscape}
\usepackage{enumitem}
\usepackage{rotating}
\usepackage{setspace}
\usepackage{threeparttable}
\usepackage{booktabs}
\usepackage[compact]{titlesec}
\usepackage{lmodern}
\fontfamily{lmtt}\selectfont
\usepackage[T1]{fontenc}

\usepackage[nocitation]{apacite}
\usepackage{natbib}
\bibpunct{(}{)}{;}{a}{}{,}
\usepackage{hyperref}
\usepackage{titling}
\usepackage{pifont}
\usepackage[capposition=top]{floatrow}
\newcommand{\subtitle}[1]{%
  \posttitle{%
    \par\end{center}
    \begin{center}\large#1\end{center}
    \vskip0.5em}%
}
\usepackage{amsthm}
\usepackage{amsmath} 
\usepackage{multirow}
\usepackage{mathrsfs}
\usepackage{graphicx}
\usepackage{float}

\renewcommand{\P}{\textrm{P}}
\newcommand{\Var}{\textrm{Var}}
\newtheorem{theorem}{Theorem}[section]
\newtheorem{lemma}[theorem]{Lemma}

\graphicspath{ {./graphics/} }

\usepackage{mathrsfs}

\usepackage{color}

\begin{document}
\pagestyle{plain}

\newtheoremstyle{mystyle}
{\topsep}
{\topsep}
{\it}
{}
{\bf}
{.}
{.5em}
{}
\theoremstyle{mystyle}
\newtheorem{assumptionex}{Assumption}
\newenvironment{assumption}
  {\pushQED{\qed}\renewcommand{\qedsymbol}{}\assumptionex}
  {\popQED\endassumptionex}
\newtheorem{assumptionexp}{Assumption}
\newenvironment{assumptionp}
  {\pushQED{\qed}\renewcommand{\qedsymbol}{}\assumptionexp}
  {\popQED\endassumptionexp}
\renewcommand{\theassumptionexp}{\arabic{assumptionexp}$'$}

\newtheorem{assumptionexpp}{Assumption}
\newenvironment{assumptionpp}
  {\pushQED{\qed}\renewcommand{\qedsymbol}{}\assumptionexpp}
  {\popQED\endassumptionexpp}
\renewcommand{\theassumptionexpp}{\arabic{assumptionexpp}$''$}

\newtheorem{assumptionexppp}{Assumption}
\newenvironment{assumptionppp}
  {\pushQED{\qed}\renewcommand{\qedsymbol}{}\assumptionexppp}
  {\popQED\endassumptionexppp}
\renewcommand{\theassumptionexppp}{\arabic{assumptionexppp}$'''$}

\renewcommand{\arraystretch}{1.3}

\newcommand{\argmin}{\mathop{\mathrm{argmin}}}
\makeatletter
\newcommand{\grande}{\bBigg@{2.25}}
\newcommand{\enorme}{\bBigg@{5}}

\newcommand{\blind}{0}

\newcommand{\tit}{One-step weighting to generalize and transport treatment effect estimates to a target population}

\if0\blind
{{\title{\tit\thanks{We thank \textcolor{black}{the Editor, Associate Editor, and two anonymous reviewers for helpful comments}. We thank Eli Ben-Michael, Larry Han, Kosuke Imai, Yige Li, Bijan Niknam, Zhu Shen, and Yi Zhang for helpful comments and suggestions. 
This work was supported through a grant from the Alfred P. Sloan Foundation (G-2020-13946) \textcolor{black}{and an award from the Patient Centered Outcomes Research
Initiative (PCORI, ME-2022C1-25648)}.
}}
\author{Ambarish Chattopadhyay\thanks{Stanford Data Science, Stanford University, 450 Jane Stanford Way Wallenberg, Stanford, CA 94305; email: \url{hsirabma@stanford.edu}.} \and Eric R. Cohn\thanks{Department of Biostatistics, Harvard University, 1 677 Huntington Ave, Boston, MA 02115; email: \url{ericcohn@g.harvard.edu}.}\and Jos\'{e} R. Zubizarreta\thanks{Departments of Health Care Policy, Biostatistics, and Statistics, Harvard University, 180 Longwood Avenue, Office 307-D, Boston, MA 02115; email: \url{zubizarreta@hcp.med.harvard.edu}.}
}}

\date{} 

\maketitle
}\fi

\if1\blind
\title{\tit}
\date{} 
\maketitle
\fi

\begin{abstract}
\input{onestep_sec0.tex}
\end{abstract}


\begin{center}
\end{center}
\clearpage
\doublespacing

\singlespacing
\pagebreak
\pagebreak
\doublespacing

\input{onestep_sec1_ec.tex}

\input{onestep_sec2.tex}

\input{onestep_sec3_ec.tex}

\input{onestep_sec4_ec.tex}

\input{onestep_sec5_ec.tex}

\input{onestep_sec6.tex}

\input{onestep_sec7.tex}

\pagebreak

\bibliographystyle{asa}
\bibliography{bib2022}

\pagebreak
\section{Web Appendix}
\small
\input{onestep_supp.tex}

\end{document}

%% file: onestep_sec0.tex
The problem of generalization and transportation of treatment effect estimates from a study sample to a target population is central to empirical research and statistical methodology. In both randomized experiments and observational studies, weighting methods are often used with this objective. Traditional methods construct the weights by separately modeling the treatment assignment and study selection probabilities and then multiplying functions (e.g., inverses) of their estimates. In this work, we provide a justification and an implementation for weighting in a single step. We show a formal connection between this one-step method and inverse probability and inverse odds weighting. We demonstrate that the resulting estimator for the target average treatment effect is consistent, asymptotically Normal, multiply robust, and semiparametrically efficient. We evaluate the performance of the one-step estimator in a simulation study. We illustrate its use in a case study on the effects of physician racial diversity on preventive healthcare utilization among Black men in California. We provide \texttt{R} code implementing the methodology.

%% file: onestep_sec1_ec.tex
\section{Introduction}
\label{sec_introduction}

From the medical and social sciences to business and government research, the value of causal inferences from randomized experiments and observational studies depends on how well they generalize or transport to a target population of interest.
If treatment effects are modified by covariates whose distributions differ between the study sample and the target population, then effect estimates based on the study may not apply to the target \citep{degtiar2023review}.
Weighting methods show promise for generalizing and transporting treatment effect estimates from a study sample to a target population. 
Traditionally, these methods construct the weights by separately modeling and estimating unit-specific treatment assignment and study selection probabilities and then multiplying functions (e.g., inverses) of the estimated probabilities (\textcolor{black}{see,} e.g., \citealt{stuart2011use}, \citealt{tipton2018review}, \citealt{dahabreh2019generalizing}).  
If these models are correctly specified, then the weighted treatment and control groups will be balanced in expectation relative to the target population and the resulting weighted estimators will be unbiased for the target average treatment effect.

In practice, however, due to model misspecification, small samples, or sparse covariates, imbalances and biases can still exist. 
Additionally, as discussed by \cite{josey2021transporting} and \cite{lu2021you}, when the estimated treatment assignment or study selection probabilities are small, inverting or otherwise transforming them to obtain weights can magnify small prediction errors. 
The multiplicative structure of the two-step weights can further exacerbate this issue, resulting in highly variable and at times extreme weights where a few observations dominate the analysis.
In turn, this can produce unstable treatment effect estimators \citep{kang2007demystifying}.


These concerns motivate the use of one-step weights for generalization and transportation; however, to our knowledge, the causal inference literature has yet to formalize and illustrate doing so.
In this paper, we fill this methodological and practical gap, and we justify, analyze, \textcolor{black}{and evaluate} a one-step weighting method for the generalization and transportation of effect estimates in both randomized experiments and observational studies, providing new insights into weighting methodology and its applications.
Building on existing covariate balancing approaches for causal inference, 
this method directly weights the treatment and control groups towards the target population in a single step, bypassing the need to estimate treatment and study selection probabilities separately and thereby avoiding a multiplicative structure of the weights.
\textcolor{black}{Our method connects to recent calibration approaches by \cite{josey2021transporting}, \cite{li2021note} and \cite{lee2021improving}, while allowing for more flexible forms of covariate balance and accommodating more varied target populations.}
By balancing general transformations of the covariates, \textcolor{black}{the one-step method} does not require individual-level data from the target population, which may be unavailable for confidentiality reasons. \textcolor{black}{Instead}, summary statistics from the target are sufficient to execute the proposed method, which can be crucial in practice in settings where data access is limited.

Our paper proceeds as follows.
Section \ref{sec_setup} defines the notation and presents a framework for the generalization and transportation of causal inferences where the study and target populations are characterized in terms of generic probability distributions.
This framework is general and can encompass a number of generalization and transportation settings, including those  where the study population is nested within the target, those where study and target populations are disjoint, and those where the study and target populations partially overlap.\footnote{\textcolor{black}{As an example of the first case, consider the generalization of estimates from a randomized trial nested within a cohort study to the entire cohort; and as an example of the second, consider data from a randomized trial conducted in one state used to estimate the effects of treatment in another state.}}
Motivated by this framework, 
Section \ref{sec_methods} proposes a method for estimating, in a single step, weights for  generalization and transportation by targeting covariate balance while also avoiding highly-variable weights.
Section \ref{sec_asymptotic} derives the formal properties of this method, showing that, under regularity conditions, it yields a weighted estimator of the target average treatment effect that is consistent, asymptotically Normal, multiply robust, and semiparametrically efficient.
Section \ref{sec_simulation} assesses the empirical performance of the method in a simulation study and reveals how multiplicative weights, which are ubiquitous in practice, can underperform in finite samples.
\textcolor{black}{Section} \ref{sec_empirical} illustrates the method \textcolor{black}{in a case study on the effects of physician racial diversity on preventive healthcare utilization among Black men in California}.
Section \ref{sec_remarks} concludes with some ideas for future directions.

%% file: onestep_sec2.tex
\section{Framework}
\label{sec_setup}

\subsection{Setup and estimands}
\label{sec_notation}

Consider a randomized experiment or observational study of $n$ units drawn randomly from a study population $\mathcal{P}$.
Denote the target population by $\mathcal{T}$.
\textcolor{black}{Let $Y(1)$ and $Y(0)$ denote the potential outcomes under treatment and control, respectively (\citealt{neyman1923application}, \citealt{rubin1974estimating}).
 Also, let $\bm{X} \in \mathbb{R}^d$ denote the covariates, and, for the study population, let $Z$ be the treatment assignment indicator (with $Z = 1$ for treatment and $Z = 0$ for control), so that $Y^{\mathrm{obs}}: = Z Y(1) + (1-Z) Y(0)$ is the observed outcome.}

We formalize the concepts of generalization and transportation by defining and comparing the distributions that characterize the study and target populations. 
Let $\mathbb{P}$ and $\mathbb{T}$ denote the probability measures that characterize $\mathcal{P}$ and $\mathcal{T}$, with densities $p(\cdot)$ and $t(\cdot)$, respectively.
Also, let $\mathcal{P}_1$ and $\mathcal{P}_0$ be the study subpopulations of treated and control units, respectively, with corresponding measures $\mathbb{P}_1  = \mathbb{P}_{\cdot|Z=1} $ and $\mathbb{P}_0  = \mathbb{P}_{\cdot|Z=0}$ and densities $p_1(\cdot)$ and $p_0(\cdot)$.

The estimand of primary interest is the target average treatment effect (\citealt{kern2016assessing}),
\begin{equation}\label{TATE1}
\tau := E_{\mathbb{T}}\{Y(1) - Y(0)\}
= \int \{y(1) - y(0)\} d\mathbb{T}_{\{Y(1),Y(0)\}}\{y(1),y(0)\}.
\end{equation}

This \textcolor{black}{expression} generalizes several \textcolor{black}{other estimands} in causal inference, including the population average treatment effect and the population average treatment effect on the treated, for certain choices of $\mathcal{T}$.
Table \ref{tab:estimands} provides examples of some of these estimands and their equivalent representations as target average treatment effects.
We note that the weighted average treatment effect\textcolor{black}{, which is defined by a tilting function $h(\bm{x})$,} equals the population average treatment effect when $h(\bm{x}) = 1$; the population average treatment effect for the treated when $h(\bm{x}) = e(\bm{x})$, where $e(\bm{x}) = \mathrm{pr}_{\mathbb{P}}(Z=1|\bm{X} = \bm{x})$ is the propensity score; and the estimand targeted by the overlap weights (\citealt{li2018balancing}) when $h(\bm{x}) = e(\bm{x})\{1-e(\bm{x})\}$.
\begin{table}[H]
\caption{\label{tab:estimands} Equivalence between $\tau$ and other causal estimands.}
\centering
\scalebox{0.87}{
\begin{tabular}{p{0.25\linewidth}cc} 
\hline
Estimand & Definition & Target \textcolor{black}{D}istribution\\ \hline
Population Average \newline Treatment Effect & $ E_{\mathbb{P}}\{Y(1) - Y(0)\}$ & $\mathbb{T}_{\{Y(1),Y(0),\bm{X}\}} = \mathbb{P}_{\{Y(1),Y(0),\bm{X}\}}$\\
        Population Average \newline Treatment Effect \newline for the Treated & $E_{\mathbb{P}}\{Y(1) - Y(0)|Z=1\}$ & $\mathbb{T}_{\{Y(1),Y(0),\bm{X}\}} = \mathbb{P}_{1, \{ {Y(1),Y(0),\bm{X}}\}}$\\
Population Average \newline Treatment Effect \newline for the Controls & $E_{\mathbb{P}}\{Y(1) - Y(0)|Z=0\}$ & $\mathbb{T}_{\{Y(1),Y(0),\bm{X}\}} = \mathbb{P}_{0,\{{Y(1),Y(0),\bm{X}}\}}$\\
Weighted Average \newline Treatment Effect & $\frac{E_{\mathbb{P}}[h(\bm{X})\{Y(1) - Y(0)\}]}{E_{\mathbb{P}}\{h(\bm{X})\}}$ & $t(y(1),y(0),\bm{x}) = p\{y(1),y(0),\bm{x}\}\frac{h(\bm{x})}{E_{\mathbb{P}}\{h(\bm{X})\}}$\\
\hline
\end{tabular}
}
\end{table}

\subsection{Assumptions and identification}
\label{sec_gen_ident}

The target average treatment effect $\tau$ can be nonparametrically identified in both randomized experiments and observational studies under the following conditions.
\begin{assumption}
\label{assump1}
\normalfont
\begin{enumerate}[label=(\alph*),leftmargin=*]
\item[]
    \item \textit{Common support of $\bm{X}$ across treatment groups:} $\text{Supp}(\mathbb{P}_{1,\bm{X}})$ =  $\text{Supp}(\mathbb{P}_{0,\bm{X}})$, i.e., $0<\mathrm{pr}_{\mathbb{P}}(Z=1|\bm{X} = \bm{x})<1$ for all $\bm{x} \in \text{Supp}(\mathbb{P}_{\bm{X}})$.

    \item \textit{Conditional exchangeability across treatment groups given $\bm{X}$:} $\mathbb{P}_{1,\{Y(1),Y(0)\}|\bm{X}} = \mathbb{P}_{0,\{Y(1),Y(0)\}|\bm{X}}$. In randomized experiments, this is achieved marginally and by design.
    
    \item \textit{Common support of $\bm{X}$ across the study and target populations: } $\text{Supp}(\mathbb{P}_{\bm{X}})$ =  $\text{Supp}(\mathbb{T}_{\bm{X}})$.
    
    \item \textit{Conditional exchangeability across the study and target populations given $\bm{X}$: } $\mathbb{T}_{\{Y(1),Y(0)\}|\bm{X}} = \mathbb{P}_{\{Y(1),Y(0)\}|\bm{X}}$.
\end{enumerate}
\end{assumption}
Assumptions 1(a) and 1(b) refer to the \textcolor{black}{overlap} and unconfoundedness assumptions in observational studies, respectively \citep{rosenbaum1983central}. 
In randomized experiments, these two assumptions hold by design. 
Assumptions 1(c) and 1(d) are their analogs for generalization and transportation. 
Assumptions 1(a) and 1(c) jointly imply that $\bm{X}$ has common support across the two treatment groups and the target population.
Similarly, Assumptions 1(b) and 1(d) jointly imply that the conditional distributions of the potential outcomes given the covariates are exchangeable across the treatment groups and the target population. 
\textcolor{black}{Since these assumptions involve unobservable potential outcomes, they cannot be directly tested; however, there are approaches to check some of their implications and assess their plausibility.
For example, with access to adequate data, \citeauthor{imbens2015causal} (2015, Chapter 21) propose  tests on pseudo-outcomes, which are known not to be affected by treatment, or pseudo-treatments, which are known not to affect outcomes.
The assumptions also become more plausible when $\bm{X}$ includes many covariates and/or when subject matter knowledge is used to motivate how the variables in $\bm{X}$ relate to the treatment assignment or study selection mechanisms.}
See the Supplementary Materials for more discussion on these conditions.

Under Assumption \ref{assump1}, $\tau$ can be identified and estimated by weighting.
For each treatment group, these weights are constructed by taking the ratio of the joint density of the covariates in the target population to their joint density in the treatment group, akin to importance sampling.
More formally, by Assumption \ref{assump1}, 
\begin{align}
    \tau = \frac{E_{\mathbb{P}_1}\left\{\frac{t(\bm{X})}{e(\bm{X})p(\bm{X})}Y^{\mathrm{obs}}\right\}}{E_{\mathbb{P}_1}\left\{\frac{t(\bm{X})}{e(\bm{X})p(\bm{X})}\right\}} - \frac{E_{\mathbb{P}_0}\left[\frac{t(\bm{X})}{\{1-e(\bm{X})\}p(\bm{X})}Y^{\mathrm{obs}}\right]}{E_{\mathbb{P}_0}\left[\frac{t(\bm{X})}{\{1-e(\bm{X})\}p(\bm{X})}\right]}, \label{identify_weighting}
\end{align}
where $e(\bm{x}) := \mathrm{pr}_{\mathbb{P}}(Z=1|\bm{X} = \bm{x})$ is the propensity score. 
This suggests using a H\'{a}jek estimator for $\tau$ by replacing expectations with sample means and $t(\bm{X})$, $e(\bm{X})$, and $p(\bm{X})$ with estimates $\widehat{t}(\bm{X})$, $\widehat{e}(\bm{X})$, and $\widehat{p}(\bm{X})$.

Our framework covers the widely analyzed nested study design setting, where
the study population $\mathcal{P}$ is nested within a larger population $\mathcal{Q}$.
For generalization, the target population is this larger population (i.e., $
\mathcal{T} = \mathcal{Q}$), and for transportation, the target population is the part of the larger population that is not included in the study (i.e., $\mathcal{T} = \mathcal{Q} \backslash \mathcal{P}$). 
In the former case, identification equation \eqref{identify_weighting} reduces to the inverse probability weighting representation for generalization (\citealt{dahabreh2019generalizing}),
\begin{align}
     \tau = \frac{E_{\mathbb{P}_1}\left\{\frac{1}{e(\bm{X})\pi(\bm{X})}Y^{\mathrm{obs}}\right\}}{E_{\mathbb{P}_1}\left\{\frac{1}{e(\bm{X})\pi(\bm{X})}\right\}} - \frac{E_{\mathbb{P}_0}\left[\frac{1}{\{1-e(\bm{X})\}\pi(\bm{X})}Y^{\mathrm{obs}}\right]}{E_{\mathbb{P}_0}\left[\frac{1}{\{1-e(\bm{X})\}\pi(\bm{X})}\right]},
    \label{eq:ipw_identify2}
\end{align}
where $\pi(\bm{x}) = \mathrm{pr}_{\mathbb{Q}}(D = 1 | \bm{X} = \bm{x})$ and $D$ is the indicator of selection into the study. Similarly, in the latter case, identification equation \eqref{identify_weighting} reduces to the inverse odds weighting representation for transportation \citep{westreich2017transportability},
\begin{align}
     \tau =  \frac{E_{\mathbb{P}_1}\left\{\frac{1-\pi(\bm{X})}{e(\bm{X})\pi(\bm{X})}Y^{\mathrm{obs}}\right\}}{E_{\mathbb{P}_1}\left\{\frac{1-\pi(\bm{X})}{e(\bm{X})\pi(\bm{X})}\right\}} - \frac{E_{\mathbb{P}_0}\left[\frac{1-\pi(\bm{X})}{\{1-e(\bm{X})\}\pi(\bm{X})}Y^{\mathrm{obs}}\right]}{E_{\mathbb{P}_0}\left[\frac{1-\pi(\bm{X})}{\{1-e(\bm{X})\}\pi(\bm{X})}\right]}.
    \label{eq:iow_identify2}
\end{align}

%% file: onestep_sec3_ec.tex
\section{Methods}
\label{sec_methods}

\subsection{Weighting in one step}
\label{sec_onego}

\textcolor{black}{In} both inverse probability and inverse odds weighting, one often fits separate models for $e(\bm{x})$ and $\pi(\bm{x})$ and then multiplies the inverses (or functions) of the estimated probabilities to obtain the final weights. 
As discussed in Section \ref{sec_introduction}, such weights can be highly variable and fail to balance the covariates relative to the target population, producing biased and unstable estimators.
To address these issues, we formalize and use a weighting method that balances each treatment group toward the target in one step. 

For $z \in (0,1)$, let $m_z(\bm{x}) := E_{\mathbb{P}}\{Y(z)|\bm{X}=\bm{x}\}$ be the conditional mean function of $Y(z)$ given $\bm{X} = \bm{x}$. By Assumption \ref{assump1}(d), $m_z(\bm{x}) = E_{\mathbb{T}}\{Y(z)|\bm{X}=\bm{x}\}$, so $E_{\mathbb{T}}\{Y(z)\} = E_{\mathbb{T}}\{m_z(\bm{X})\}$. By Assumption \ref{assump1}(b), for normalized weights $w_i$, the bias of the H\'{a}jek estimator for $E_{\mathbb{T}}\{Y(z)\}$ is,
\begin{align} 
E_{\mathbb{P}}\left(\sum_{i:Z_i=z}w_i Y^{\mathrm{obs}}_i\right) - E_{\mathbb{T}}\left\{Y(z)\right\}  &=  E_{\mathbb{P}}\left[ \sum_{i:Z_i=z}w_i m_z(\bm{X}_i) - E_{\mathbb{T}}\{m_z(\bm{X})\}  \right].\label{eq:bias1}
\end{align}
In other words, weights that exactly balance the mean of $m_z(\bm{x})$ in treatment group $Z = z$ relative to $E_{\mathbb{T}}\{m_z(\bm{X})\}$ \textcolor{black}{(i.e., the mean of $m_z(\bm{x})$ in treatment group $Z = z$ after weighting becomes equal to $E_{\mathbb{T}}\{m_z(\bm{X})\}$)} completely remove the bias of $\sum_{i:Z_i=z}w_i Y^{\mathrm{obs}}_i$ as an estimator of $E_{\mathbb{T}}\{Y(z)\}$.
Assuming that $\Var_{\mathbb{P}}\left\{Y(z)|\bm{X} = \bm{x}\right\} = \sigma^2_z$ for $z \in (0,1)$, for all $\bm{x} \in \mathcal{X}$, we can compute the conditional variance of the weighted estimator as
\begin{equation}\label{eq:var1}
    \Var_{\mathbb{P}}\left(\sum_{i:Z_i=z}w_i Y^{\mathrm{obs}}_i \Big| \bm{X},\bm{Z} \right) = \sigma^2_z  \sum_{i:Z_i=z}w^2_i.
\end{equation}
This says that in order to obtain a stable H\'{a}jek estimator, it is crucial to have a set of weights with small \textcolor{black}{squared or} $L_2$ norm.  
Together, Equations \ref{eq:bias1} and \ref{eq:var1} suggest the use of weights of \textcolor{black}{minimal} $L_2$ norm that balance the conditional mean functions $m_1(\bm{x})$ and $m_0(\bm{x})$ towards the target population of interest, and this guides our method.

\subsection{Balancing towards the target population}
\label{sec_balancing}

Building on the minimal weights of \cite{wang2020minimal}, we solve the following convex optimization problem,
\begin{align}\label{minimal}
\argmin_{\bm{w}} \left\{  \sum_{i: Z_i=z}\psi(w_i) : \left|\sum_{i: Z_i=z} w_i B_k(\bm{X}_i) - \bar{B}^*_k    \right| \leq \delta_k, \; k = 1, 2, ..., K; \sum_{i:Z_i=z}w_i = 1
\right\},
\end{align}
where $B_k(\bm{X})$, $k = 1, 2, ..., K$ is a set of basis functions that spans $m_z(\bm{x})$, as motivated by Equation \ref{eq:bias1}, and $\psi$ is a smooth measure of the dispersion of the weights, such as their $L_2$ norm, as motivated by Equation \ref{eq:var1}.

The solution to \eqref{minimal} is the vector of weights $\bm{w}$ of minimum dispersion that approximately balances the weighted mean of the basis functions $B_k(\bm{X})$ in treatment group $Z = z$ relative to its mean in the target population, $\bar{B}_k^*$ \textcolor{black}{(i.e., the mean of $B_k(\bm{X})$ in treatment group $Z = z$ becomes approximately equal to $\bar{B}_k^*$ after weighting)}.
\textcolor{black}{The vector $ (\bar{B}_1^*,...,\bar{B}_K^*)^\top$ is the covariate profile, and access to this summary measure from the target population is sufficient to execute the method.}
The basis functions are flexible, ranging from main terms (which is sufficient for removing bias if $m_z(\bm{x})$ is linear in the covariates) to higher-order interactions (sufficient if $m_z(\bm{x})$ is linear in these interactions) to the basis functions corresponding to a particular kernel (sufficient if $m_z(\bm{x})$ lies in a reproducing kernel Hilbert space, RKHS).
See \cite{ben2021balancing} for a more detailed discussion.
Additional constraints in \eqref{minimal} can also force the weights to be non-negative and produce a sample-bounded estimator \citep{robins2007comment}.
Finally, the parameters $\delta_k$, $k = 1, ..., K$, are set by the investigator, ultimately trading off bias for variance (see \citealt{chattopadhyay2020balancing} for a discussion and guidance). 
\textcolor{black}{Following \cite{chattopadhyay2023implied} and \cite{bruns2023augmented}, in the Supplementary Materials, we show how, for certain choices of the parameters in \eqref{minimal}, the one-step weighting estimator is equivalent to the linear regression imputation estimators for generalization and transportation (\citealt{dahabreh2019generalizing}, \citealt{dahabreh2020extending}).}

In this way, the treatment groups are directly balanced relative to the target population in one step with weights of minimal dispersion that produce a stable estimator. This method is applicable to various generalization and transportation problems, including those where the study and target populations overlap and those where they are disjoint.

\textcolor{black}{We note that when the estimand is a weighted average treatment effect (WATE; see Table \ref{tab:estimands}), with standard choices of $h(\cdot)$ (e.g., $h(\bm{x}) = 1$ or $h(\bm{x}) = e(\bm{x})\{1-e(\bm{x})\}$), the two-step weighting method boils down to a one-step method that requires estimation of the propensity score only. With a correctly specified model for the propensity score, this method is expected to perform similarly to the proposed one-step method. Under model misspecification, the proposed one-step method may improve estimation efficiency since it explicitly controls the variability of the weights, while directly targeting covariate balance. 
}

%% file: onestep_sec4_ec.tex
\section{Asymptotic properties}
\label{sec_asymptotic}

Implicitly, the one-step weighting method simultaneously models both the true inverse probability (or inverse odds) weights as well as the potential outcomes, the former via the dual optimization problem of \eqref{minimal} and the latter via the balancing conditions in \eqref{minimal}. 
In fact, since the one-step weights are obtained by solving \eqref{minimal} for the treated and control units separately, this method posits two separate models for the inverse probability weights of the treated and control units, and two separate models for the potential outcomes under treatment and control. These models are nonparametric in the sense that they uniformly approximate the true (weight or outcome) functions using a set of basis functions that grows with the sample size. \textcolor{black}{These implicit models allow us to derive the asymptotic properties of the one-step weights and estimator.}

{\color{black}
In this section, we provide asymptotic results for the generalization setting where the study sample is nested within a random sample of size $n^*$ from the target population. 
In this setting, Assumption \ref{assump2} states that the true inverse probability weights can be uniformly approximated by the basis functions $\bm{B}(\cdot)$ and Assumption \ref{assump3} states a similar result for the conditional mean functions of the potential outcomes.
To derive our theoretical results, we let the number of basis functions $K$ grow with $n^*$, akin to nonparametric sieve estimation (\citealt{andrews1991asymptotic}, \citealt{newey1997convergence}; see also \citealt{fan2016improving}).

\begin{assumption}
\label{assump2}
\normalfont
The true inverse probability weights for group $Z=z$ satisfy $w^{\text{IP}}(\bm{x}) = n^* \rho'\{g^{*}_z(\bm{x})\}$ for all $\bm{x} \in \mathcal{\bm{X}}$, where  $g^*_z(\cdot)$ is a smooth function such that $\sup_{\bm{x} \in \bm{X}}|g^*_z(\bm{x}) - \bm{B}(\bm{x})^\top \bm{\lambda}^*_{1z}| = O(K^{-r_z})$ for some $\bm{\lambda}^*_{1z} \in \mathbb{R}^K$ and $r_z>1$.
\end{assumption}

The function $\rho(\cdot)$ is obtained from the dual problem of \eqref{minimal} (the exact form is provided in \eqref{dual1} in the Supplementary Materials), and
$g_z^*(\cdot)$ is a smooth link function between the weights and the basis functions.
Assumption \ref{assump2} is satisfied in the parametric case where $g_z^*(\cdot)$ is a linear combination of a finite number of basis functions.

\begin{assumption}\normalfont
The conditional mean function of the potential outcome under treatment $Z=z$ satisfies $\sup_{\bm{x}\in \mathcal{X}}|m_z(\bm{x}) - \bm{B}(\bm{x})^\top \bm{\lambda}^*_{2z}| = O(K^{-s_z}) $ for some $\bm{\lambda}^*_{2z} \in \mathbb{R}^K$ and $s_z>3/4$, with $||\bm{\lambda}^*_{2z}||_2 ||\bm{\delta}||_2 = o(1)$.
\label{assump3}
\end{assumption}

Assumption \ref{assump3} is satisfied when $m_z(\cdot)$ is a linear combination of a finite number of basis functions.
The condition $||\bm{\lambda}^*_{2z}||_2 ||\bm{\delta}||_2 = o(1)$ is a technical requirement that ensures that the coefficients of $\bm{B}(\cdot)$ do not dominate the overall degree of approximate balance as the number of basis functions goes to infinity.

Leveraging these two assumptions and the regularity conditions in Assumption \ref{assump4} below, we can establish multiple robustness of the one-step H\'{a}jek estimator of the target average treatment effect.
At a high level, these regularity conditions pertain to the dual objective function and basis functions that are used to approximate the true weights and the potential outcomes.
They are similar to the technical conditions in Assumption 1 of \cite{wang2020minimal} and Assumption 4.1 in \cite{fan2016improving}. See Appendix \ref{sec_appendix_convergence} for details.

\begin{assumption} \normalfont
\label{assump4}
For $z \in \{0,1\}$, 
\begin{enumerate}[label=(\alph*)]

\item There exist constants $c_0$, $c_1$, $c_2$ with $0<c_0<1/2$ and $c_1<c_2<0$, such that $c_1 \leq n^*\rho{''}(v) \leq c_2$ for all $v$ in a neighborhood of $\bm{B}(\bm{x})^\top \bm{\lambda}^*_{1z}$. Also, $c_0 \leq 1/(n^*\rho'(v)) \leq 1-c_0$ for all $v = \bm{B}(\bm{x})^\top \bm{\lambda}$, $\bm{x} \in \mathcal{X}$, $\bm{\lambda}$. 
    
    \item $\sup\limits_{x \in \mathcal{X}}||\bm{B}(\bm{x})||_2 \leq CK^{1/2}$ and $||E_{\mathbb{T}}\{\bm{B}(\bm{X}) \bm{B}(\bm{X})^\top\}||_F \leq C$, for some constant $C>0$, where $||\cdot||_F$ denotes the Frobenius norm.
  
    \item $K = O\{(n^*)^\alpha\}$ for some $0<\alpha<2/3$.
    
    \item For some constant $C>0$, $\lambda_{\text{min}}\Big[ E_{\mathbb{T}}\{D\mathbbm{1}(Z = z)\bm{B}(\bm{X})\bm{B}(\bm{X})^\top\}\Big]>C$, where $\lambda_{\min}(\bm{A})$ denotes the smallest eigenvalue of $\bm{A}$.
    
    \item $||\bm{\delta}||_2 = O_P\left[K^{1/4}\left\{(\log K) / n^*\right\}^{1/2} + K^{- r_z + 1/2}\right]$.
\end{enumerate}
\end{assumption}
Assumption \ref{assump4} assigns smoothness conditions on the objective function through $\rho(\cdot)$,\footnote{\textcolor{black}{These conditions are satisfied for typical choices of convex objective functions $\psi(\cdot)$, e.g., those corresponding to entropy balancing \citep{hainmueller2012balancing} and stable balancing weights \citep{zubizarreta2015stable}.}}
ensures that the vector of basis functions $\bm{B}(\cdot)$ are sufficiently well-behaved in terms of its length and dimension, and controls the growth rate for the degree of approximate balance $\bm{\delta}$. 

Given the above assumptions, Theorem \ref{thm_consistency} establishes multiple consistency conditions for the one-step H\'{a}jek estimator of the TATE.


\begin{theorem}[Consistency] \normalfont
\label{thm_consistency}
Suppose that Assumptions \ref{assump1} and \ref{assump4} hold. Then
the one-step H\'{a}jek estimator $\hat{\tau} = \sum_{i:Z_i=1}\widehat{w}_iY^{\text{obs}}_i - \sum_{i:Z_i=0}\widehat{w}_iY^{\text{obs}}_i$ is consistent for the target average treatment effect $\tau$ if any of the following conditions are satisfied.
\begin{enumerate}[label =(\alph*)]
 \item The models for the inverse probability weights of the treated and control units are both correctly specified (Assumption \ref{assump2} holds for $z \in \{0, 1\}$).
 \item The models for the potential outcomes under treatment and control are both correctly specified (Assumption \ref{assump3} holds for $z \in \{0, 1\}$).
 \item The models for the inverse probability weights of the treated units and the potential outcomes under control are both correctly specified (Assumption \ref{assump2} holds for $z = 1$ and Assumption \ref{assump3} holds for $z = 0$).
 \item The models for the inverse probability  weights of the control units and the potential outcomes under treatment are both correctly specified (Assumption \ref{assump2} holds for $z = 0$ and Assumption \ref{assump3} holds for $z = 1$).
\end{enumerate}
\end{theorem}

Thus, in a similar way to multiply robust estimators,  $\hat{\tau}$ admits multiple conditions for consistency, where each condition requires a suitable approximation of the implied models for the inverse probability weights, the potential outcomes, or a combination of the two.

Finally, if the four aforementioned models are correctly specified, then under some additional regularity conditions, $\hat{\tau}$ is asymptotically Normal and semiparametrically efficient for the target average treatment effect $\tau$ (\citealt{robins1995semiparametric}). 
At a high level, these additional regularity conditions pertain to the true inverse probability weights and potential outcome mean functions.
They are similar to Assumption 2 in \cite{wang2020minimal} and Assumption 4.1 in \cite{fan2016improving}. See Appendix \ref{sec_appendix_Normality} in the Supplementary Materials for details.

\begin{assumption}\normalfont
\label{assump5}
For $z \in \{0,1\}$, 
\begin{enumerate}[label=(\alph*)]

\item $E_{\mathbb{T}}\{Y^2(z)\}<\infty$.

\item Let $g^*_z(\cdot) \in \mathcal{G}_z$. $\mathcal{G}_z$ satisfies $\log N_{[]}\{\epsilon, \mathcal{G}_z, L_2(P)\} \leq C_1(1/\epsilon)^{1/k_1}$ for some constants $C_1>0$ and $k_1>1/2$, where $N_{[]}\{\epsilon, \mathcal{G}_z, L_2(P)\}$ is the covering number of $\mathcal{G}_z$ by epsilon brackets.
    
\item Let $m_z(\cdot) \in \mathcal{M}_z$. $\mathcal{M}_z$ satisfies $\log N_{[]}\{\epsilon, \mathcal{M}_z, L_2(P)\} \leq C_2(1/\epsilon)^{1/k_2}$ for some constants $C_2>0$ and $k_2>1/2$, , where $N_{[]}\{\epsilon, \mathcal{M}_z, L_2(P)\}$ is the covering number of $\mathcal{M}_z$ by epsilon brackets.

\item $(n^{*})^{\{2(r_z + s_{z} - 0.5)\}^{-1}} = o(K)$, where $r_z, s_z$ are the constants in assumptions 2 and 3, respectively. 
    
\end{enumerate}
\end{assumption}

Assumption \ref{assump5} controls the complexity of the function classes $\mathcal{G}_z$ and $\mathcal{M}_z$ corresponding to the IP weights and potential outcomes, respectively, along with additional restrictions on the growth rate of the number of basis functions. 

Under these conditions, Theorem \ref{thm_normality} establishes asymptotic Normality and semiparametric efficiency of the one-step H\'{a}jek estimator of the TATE.

}

\begin{theorem}[Asymptotic normality and semiparametric efficiency]\normalfont
\label{thm_normality}
Suppose that the study sample is nested within a random sample of size $n^*$ from the target population and that Assumptions 2-5 hold. Then, as $n^*$ tends to infinity, $(n^*)^{1/2}(\widehat{\tau} - \tau)$ converges in distribution to a Normal random variable with mean 0 and variance $V^*_\tau = E_{\mathbb{T}}(\sigma^2_1(\bm{X})/ \{\pi(\bm{X})e(\bm{X})\} + {\sigma^2_0(\bm{X}) / [\pi(\bm{X})\{1-e(\bm{X})\}]} + \{m_1(\bm{X})-m_0(\bm{X}) - \tau\}^2 )$, where $\sigma^2_z(\bm{x}) = \Var_{\mathbb{T}}\{Y(z)|\bm{X}=\bm{x}\}$ for $z \in (0,1)$.
\end{theorem}
\textcolor{black}{For generalization problems in nested study settings, $V^*_\tau$ is shown to be precisely the semiparametric efficiency bound for the TATE, regardless of whether the study is randomized or observational (\citealt{li2021note}; see also \citealt{lee2021improving}). Theorem \ref{thm_normality} states that the one-step estimator achieves the semiparametric efficiency bound.}

{\color{black}
We conclude this section with a couple of remarks. 
First, the previous asymptotic properties are derived under the assumption that the potential outcomes are homoscedastic. 
However, as shown in \cite{hirshberg2021augmented}, under heteroscedasticity, the desired asymptotic properties also hold for a similar class of balancing estimators.
Theoretically, the high-level rationale is that, since the basis functions balanced by the one-step approach are allowed to grow flexibly with the sample size, in large samples, the one-step weights end up estimating the IP weights, also under heteroscedasticity. 
Empirically, we investigate the finite sample performance of the one-step estimator under heteroscedasticity in a simulation study; see Appendix \ref{sec_simu_hetro} in the Supplementary Materials for details.

Second, the asymptotic properties are derived without explicitly constraining the weights to be non-negative.
Nevertheless, with these constraints, the one-step estimator is expected to exhibit similar asymptotic behavior as that without the constraints.
Due to the overlap condition in Assumptions 1(a) and 1(c), in large samples, the solutions to \eqref{minimal} tend to be non-negative, coinciding with the solutions where \eqref{minimal} explicitly includes the non-negativity constraints.
Proposition 1 in \cite{zhao2016entropy} provides a related discussion in the context of entropy balancing.
}

%% file: onestep_sec5_ec.tex
\section{Simulation study}
\label{sec_simulation}
We evaluate the performance of the one-step weights in a simulation study for generalizing treatment effect estimates from a study to a larger cohort in which the study is nested.
We consider both randomized and observational study settings.

Our design builds on \textcolor{black}{the study by} \cite{kang2007demystifying}.
There are four independent unobserved covariates distributed as $U_1, U_2, U_3, U_4 \sim \mathcal{N}(0, 1)$, and four observed covariates generated as $X_1 = \exp(U_1/2)$, $X_2 = U_2/\{1 + \exp(U_1)\} + 10$, $X_3 = (U_1U_3/25 + 0.6)^3$, and $X_4 = (U_2 + U_4 + 20)^{2}$.
$D$ is the binary indicator for selection into the study, and $Z$ is the binary treatment indicator.
The true model for the probability of selection into the study is $\mathrm{pr}(D = 1 | \bm{U}) = \text{expit}(-U_1 + 0.5 U_2 - 0.25 U_3 - 0.1 U_4)$ so that, marginally, $\mathrm{pr}(D = 1)  = 0.5$. 
The total cohort size is 1000.
For the randomized study setting, $\mathrm{pr}(Z = 1|\bm{U}) = 0.5$, and for the observational setting, $\mathrm{pr}(Z=1|\bm{U}) = \text{expit}(U_1 + 2U_2 - 2U_3 - U_4)$. 

There is one control potential outcome model, $Y(0) = 210 + 27.4 U_1 + 13.7 U_2 + 13.7 U_3 + 13.7 U_4 + \epsilon_0$, where $\epsilon_0 \sim \mathcal{N}(0, 5^2)$, and there are three treatment potential outcome models: Model 1 is given by $Y(1) = 210 + 27.4 U_1 + 13.7 U_2 + 13.7 U_3 + 13.7 U_4 + \epsilon_1$; Model 2, by $Y(1) = 210 + 41.1 U_1 + 13.7 U_2 + 13.7 U_3 + 13.7 U_4 + \epsilon_1$; and Model 3, by $Y(1) = 210 + 41.1 U_1 + 27.4 U_2 + 27.4 U_3 + 13.7 U_4 + \epsilon_1$; where $\epsilon_1 \sim \mathcal{N}(0,5^2)$.
Thus, the target average treatment effect is $0$ in all cases, but the degree of treatment effect heterogeneity varies across the models\textcolor{black}{, which allows us to compare methods' performance under varying degrees of treatment effect heterogeneity and scenarios where the variables determining such heterogeneity are accounted for or not in the covariate adjustments.}
For Model 1, there is no heterogeneity; for Model 2, there is heterogeneity by $U_1$; and for Model 3, there is heterogeneity by $U_1$, $U_2$, and $U_3$.
Specifically, for Model 1, $E\{Y(1) - Y(0) | \bm{U}\} = 0$, for Model 2, $E\{Y(1) - Y(0) | \bm{U}\} = 13.7 U_1$, and for Model 3, $E\{Y(1) - Y(0) | \bm{U}\} = 13.7 U_1 + 13.7 U_2 + 13.7 U_3$.

We compare three versions of one-step weighting to three versions of two-step weighting.
The first one-step method balances the mean of the first observed covariate only, the second balances the means of all of the observed covariates, and the third balances the means of all of the unobserved covariates.
For the one-step methods, \textcolor{black}{we find the weights that solve \eqref{minimal}}.
A variation of the tuning algorithm in \cite{chattopadhyay2020balancing} is used to select the imbalance tolerances, where imbalances are computed relative to the target covariate profile.
\textcolor{black}{In \eqref{minimal}, we minimize the $L_2$ norm of the weights.
Our choice stems from Equation \ref{eq:var1}, which shows that the variance of the weighted estimator is directly linked to the $L_2$ norm of the weights.}

The two-step methods \textcolor{black}{fit separate models for the two unknown functions, the probability of sample selection and the probability of treatment assignment given covariates, and plug the estimated probabilities from these models into \eqref{eq:ipw_identify2} to compute the weights.}
These models are logistic regressions of the study selection indicator on covariates and additionally of the treatment indicator on covariates in the observational study setting, where the first two-step method only uses the first observed covariate in the regression, the second uses all of the observed covariates, and the third uses all of the unobserved covariates.
The third two-step and one-step methods are correctly-specified, while the others are misspecified.

Table \ref{tab:Tab2} shows the root-mean-squared errors of the H\'{a}jek estimators, based on 800 simulations.
The one-step weights outperform the two-step weights across the three outcome models in both the randomized and observational study settings. 
In the randomized study, the one-step estimators reduce the root-mean-squared error relative to the corresponding two-step estimators by $86\%$, on average. 
In the observational study, the one-step estimators achieve $70\%$ smaller root-mean-squared error relative to the corresponding two-step estimators, on average. 
In particular, the root-mean-squared errors under the correctly specified one-step methods are an order of magnitude smaller than those under the correctly specified two-step methods. \textcolor{black}{A similar pattern is observed in settings that allow for heteroscedasticity of the potential outcome models; see Appendix \ref{sec_simu_hetro} in the Supplementary Materials for details.}

\begin{table}[H]
\singlespacing
\centering
\caption{\label{tab:Tab2} Root-mean-squared error of the H\'{a}jek estimator of the target average treatment effect using different weighting methods in both the randomized and observational study settings.}
\begin{tabular}{lrrrrrrr}
                  & \multicolumn{3}{c}{Randomized Study Setting} &  & \multicolumn{3}{c}{Observational Study Setting} \\ \cline{2-4} \cline{6-8} 
                  & Outcome       & Outcome       & Outcome      &  & Outcome        & Outcome        & Outcome       \\
Weighting Method  & Model 1       & Model 2       & Model 3      &  & Model 1        & Model 2        & Model 3       \\ \hline
Two-Step 1 & 17.47         & 21.80         & 23.53        &  & 22.55          & 24.76          & 24.96         \\
One-Step 1 & 2.91          & 3.17          & 4.40         &  & 9.18           & 8.64           & 7.58          \\
Two-Step 2 & 19.00         & 23.70         & 25.23        &  & 42.47          & 41.44          & 45.72         \\
One-Step 2 & 2.26          & 2.53          & 3.27         &  & 17.32          & 16.38          & 24.16         \\
Two-Step 3 & 4.75          & 5.58          & 6.09         &  & 9.11           & 10.59          & 12.43         \\
One-Step 3 & 0.54          & 0.70          & 0.91         &  & 0.73           & 0.90           & 1.14      \\
\hline
\end{tabular}
\end{table}

\begin{centering}
\begin{figure}[H]
\centering
\caption{Effective sample sizes and maximum normalized weights across weighting methods in randomized and observational study settings.}\label{fig:ess}
\includegraphics[scale=0.9]{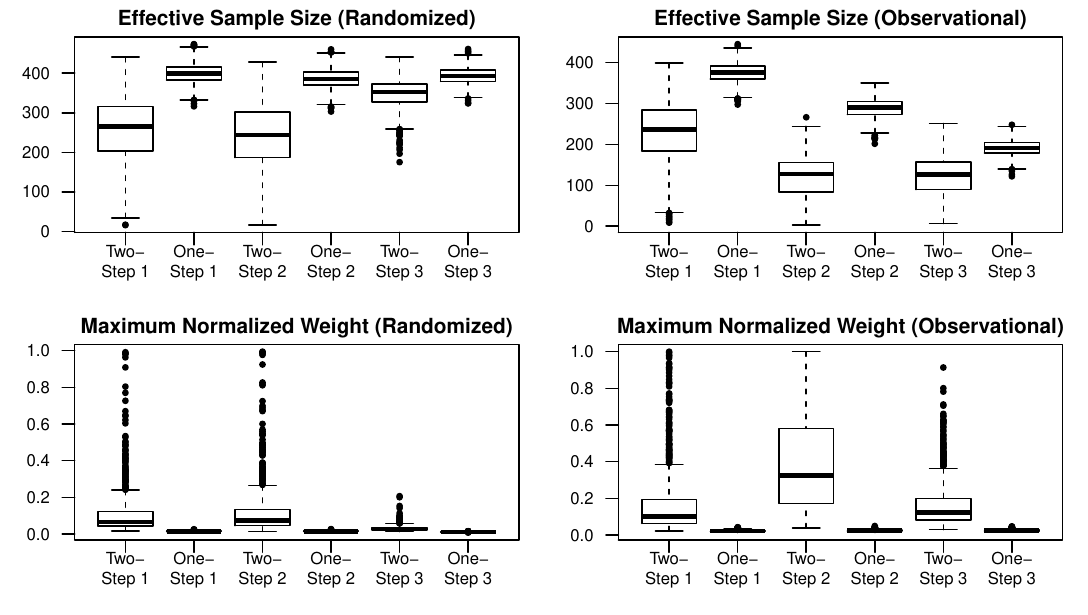}
\end{figure}
\end{centering}

These improvements are at least partially driven by the high variability and extremity of the two-step weights. 
Figure \ref{fig:ess} shows how the effective sample sizes and maximum normalized weights vary across simulations.
The effective sample size relates to the variance of the weights, where, roughly, higher variance corresponds to lower effective sample size (see, e.g., \citealt{chattopadhyay2020balancing}). Similarly, the maximum normalized weight provides a measure of the extremity of the weights.

In the randomized study, the one-step method leads to a $43\%$ improvement in the effective sample size compared to the two-step method on average.
In the observational study, the one-step method leads to an $87\%$ improvement in the effective sample size over the two-step method on average: more than double its improvement in the randomized study. 
Here, we see how the multiplicative structure of the two-step weights exacerbates the problem of extreme weights when there is less overlap across treatment groups, as is the case in the observational setting as compared to the randomized one.
Finally, in both study settings, the maximum normalized weights under the two-step method tend to be more extreme than those under the one-step method, particularly for the observational study. 
In some cases, the most extreme normalized weight under the two-step method is very close to one, signifying that nearly all information used to construct the treatment effect estimator comes from a single unit.

In the current simulation setup, the study population is contained in the target population. 
\textcolor{black}{To investigate the performance of the one-step method for other target populations, we consider another simulation study where the study and target populations are disjoint.
See Appendix \ref{appendix_simulation} in the Supplementary Materials.}
Here, too, the one-step methods outperform the two-step methods.

%% file: onestep_sec6.tex
\section{Case Study}
\label{sec_empirical}
We illustrate how the one-step weighting method can be used to generalize and transport causal effect estimates using data from an experiment on the effects of physician racial diversity on preventive healthcare utilization among Black men in California \citep{AlsanMarcella2019DDMf}.
The study was designed to test whether Black men were more or less likely to take up preventive health services if served by a Black doctor instead of a White doctor.
Out of 1,300 Black men recruited to participate, 637 presented to the clinic for screening, resulting in 313 randomized to receive care from a Black doctor and 324 from a White doctor.
After meeting with their assigned doctor for an initial consultation, participants could elect to receive any of five preventive health services (four cardiovascular screening services and a flu shot). 
There are seven outcomes: five corresponding to the selection of each particular service, one measuring the proportion of the cardiovascular services selected, and one measuring the proportion of invasive services (i.e., those requiring a blood sample or an injection) selected.

We estimate average treatment effects for three target populations of interest: a high medical mistrust population (defined as the subset of the trial population scoring high on a measure of medical mistrust), the entire recruited population, and White men (defined using the sample of White men erroneously presenting to and randomized at the clinic). 
Importantly, the validity of inferences for this last population relies on the assumption that the covariate profile is sufficient to characterize any effect modification by race.
Table \ref{tab:tab4} presents the profiles for these target populations.

\begin{centering}
\begin{table}[H]
\singlespacing
\caption{Covariate profiles for the study sample and target populations.}
\label{tab:tab4}
\scalebox{0.65}{
\begin{tabular}{lrrrr}
\hline
                                  & \multicolumn{1}{r}{}       & \multicolumn{3}{c}{Target Populations}                                                       \\ \cline{3-5} 
                                  & \multicolumn{1}{c}{Study}  & \multicolumn{1}{c}{High Medical} & \multicolumn{1}{c}{Recruited} & \multicolumn{1}{c}{White} \\
Covariate                         & \multicolumn{1}{c}{Sample} & \multicolumn{1}{c}{Mistrust}     & \multicolumn{1}{c}{Sample}    & \multicolumn{1}{c}{Men}   \\ \hline
ER visits                         & 1.6                        & 1.9                              & 1.4                           & 1.8                       \\
\quad Missing (\%) & 19.8                       & 26.3                             & 19.8                          & 25.0                      \\
Nights in hospital                & 1.3                        & 1.3                              & 1.4                           & 0.1                       \\
\quad Missing (\%)  & 19.8                       & 28.3                             & 19.1                          & 25.0                      \\
Hospital visits                   & 4.4                        & 3.3                              & 4.3                           & \multicolumn{1}{r}{NA}    \\
\quad Missing (\%)  & 29.4                       & 33.3                             & 27.3                          & \multicolumn{1}{r}{NA}    \\
Age                               & 44.5                       & 43.5                             & 42.8                          & 42.5                      \\
\quad Missing (\%)  & 2.7                        & 2.0                              & 3.5                           & 8.3                       \\
Self-rated health (\%)             &                            &                                  &                               &                           \\
\quad Good         & 60.6                       & 56.6                             & 67.0                          & 75.0                      \\
\quad Not good     & 27.8                       & 32.3                             & 22.3                          & 8.3                       \\
\quad Missing      & 11.6                       & 11.1                             & 10.7                          & 16.7                      \\
Health problems (\%)               &                            &                                  &                               &                           \\
\quad Yes          & 58.6                       & 62.6                             & 57.0                          & 58.3                      \\
\quad No           & 37.8                       & 36.4                             & 39.5                          & 41.7                      \\
\quad Missing      & 3.6                        & 1.0                              & 3.5                           & 0.0                       \\
Medical mistrust (\%)              &                            &                                  &                               &                           \\
\quad Low          & 51.2                       & 0.0                              & 50.5                          & 50.0                      \\
\quad Medium       & 29.2                       & 0.0                              & 29.8                          & 25.0                      \\
\quad High         & 15.5                       & 100.0                            & 15.5                          & 25.0                      \\
\quad Missing      & 4.1                        & 0.0                              & 4.2                           & 0.0                       \\
Has PCP (\%)                       &                            &                                  &                               &                           \\
\quad Yes          & 52.1                       & 31.3                             & 55.8                          & 25.0                      \\
\quad No           & 32.2                       & 40.4                             & 29.4                          & 41.7                      \\
\quad Missing      & 15.7                       & 28.3                             & 14.8                          & 33.3                      \\
Uninsured (\%)                     &                            &                                  &                               &                           \\
\quad Yes          & 22.8                       & 24.2                             & 21.8                          & 50.0                      \\
\quad No           & 58.4                       & 50.5                             & 61.7                          & 33.3                      \\
\quad Missing      & 18.8                       & 25.3                             & 16.5                          & 16.7   \\ 
Married (\%)                      &                            &                                  &                               &                           \\
\quad Yes          & 13.8                       & 14.1                             & 16.6                          & 33.3                      \\
\quad No           & 78.2                       & 77.8                             & 76.8                          & 58.3                      \\
\quad Missing      & 8.0                        & 8.1                              & 6.6                           & 8.3                       \\
High school education or less (\%) &                            &                                  &                               &                           \\
\quad Yes          & 54.9                       & 51.5                             & 47.2                          & 50.0                      \\
\quad No           & 32.3                       & 34.3                             & 41.5                          & 41.7                      \\
\quad Missing      & 12.7                       & 14.1                             & 11.3                          & 8.3                       \\
Low income (\%)                    &                            &                                  &                               &                           \\
\quad Yes          & 40.2                       & 50.5                             & 31.6                          & 25.0                      \\
\quad No           & 49.5                       & 38.4                             & 59.5                          & 75.0                      \\
\quad Missing      & 10.4                       & 11.1                             & 8.9                           & 0.0                       \\
Public benefits (\%)               &                            &                                  &                               &                           \\
\quad Yes          & 26.7                       & 14.1                             & 21.8                          & \multicolumn{1}{r}{NA}    \\
\quad No           & 64.8                       & 77.8                             & 71.4                          & \multicolumn{1}{r}{NA}    \\
\quad Missing      & 8.5                        & 8.1                              & 6.8                           & \multicolumn{1}{r}{NA}\\   
\hline
\end{tabular}
}
\begin{flushleft}
\scriptsize{
\vspace{15pt}
The White sample that was erroneously randomized was not asked questions about hospital visits and public benefits and hence this profile is ``NA'' for these covariates. NA = not applicable, ER = emergency room, and PCP = primary care provider.}
\end{flushleft}
\end{table}
\end{centering}

\begin{centering}
\begin{figure}[H]
\singlespacing
\caption{Distributions of target absolute standardized mean differences and effective sample sizes for three target populations.}\label{fig:fig1}
\includegraphics[scale=0.7]{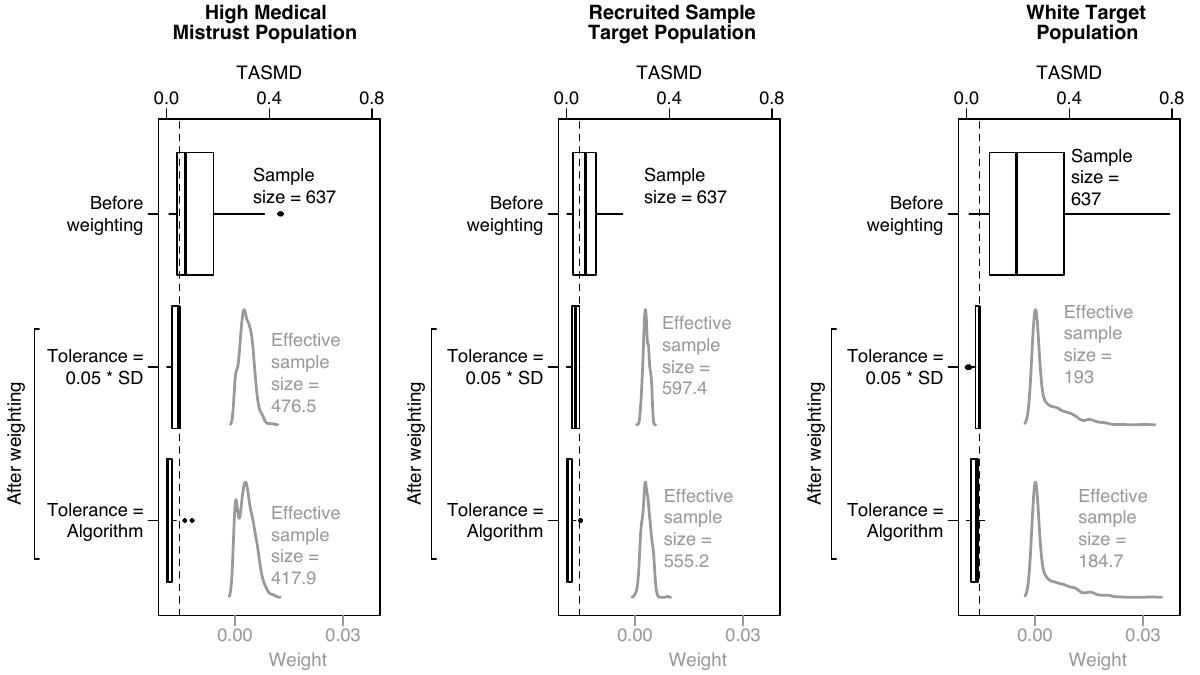}
\begin{flushleft}
\scriptsize{TASMD = target absolute standardized mean difference. The black vertical dashed line in each plot marks a TASMD of 0.05, signifying the heuristic that a TASMD $<$ 0.05 indicates good balance. SD = standard deviation.}
\end{flushleft}
\end{figure}
\end{centering}

Figure \ref{fig:fig1} summarizes the performance of two implementations of the one-step weighting method for achieving balance relative to the various target covariate profiles. In all implementations, weights were constrained to be non-negative to avoid extrapolation beyond the support of the observed data. One implementation of the method restricts the tolerated post-weighting covariate imbalances to be no higher than a constant factor (i.e., 0.05) times each covariate's standard deviation in the target, and the other implementation uses a variation of the tuning algorithm in \cite{chattopadhyay2020balancing} to determine these tolerances. Balance is evaluated using the target absolute standardized mean difference, which measures the standardized difference in means of a covariate between the study and target samples. The figures also summarize the dispersion of the weights via density plots and effective sample sizes. Lower variability of the weights translates into a higher effective sample size. Thus, we see that, as profiles of the target populations differ more from the covariate means of the trial sample, dispersion increases and the effective sample size decreases. However, these effective sample sizes are optimal in the sense that they are the maximum achievable effective sample sizes subject to the covariate balance requirements relative to the target.

\begin{figure}[H]
\singlespacing
\centering
\caption{Estimates of the target average treatment effect for various outcome variables and target populations.}\label{fig:fig2}
\includegraphics[scale=0.75]{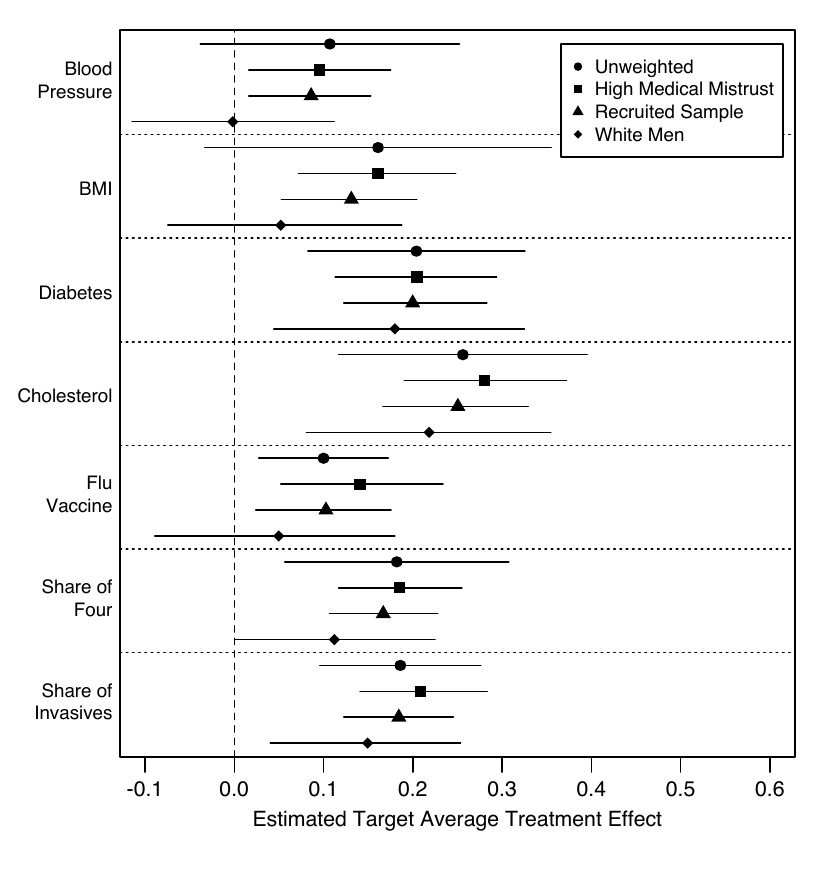}
\end{figure}

We evaluate the seven outcomes analyzed in the original study. Like other design-based methods for covariate adjustment, we can use the same weights to estimate treatment effects for all outcomes, for a given target population. Figure \ref{fig:fig2} presents the 
H\'{a}jek estimates of the target average treatment effect for each outcome using the original unweighted sample and the weighted samples corresponding to each target population, along with bootstrapped confidence intervals, whose properties for weighted estimators have been discussed in related works \citep{austin2016variance, zhao2019sensitivity}. 
Overall, weighted estimates for the various target populations are fairly consistent with each other and with the unweighted results.
The high medical mistrust population shows slightly larger treatment effects for some outcomes, particularly for the flu shot.
Additionally, estimated effects for the white target population are slightly smaller and/or are more likely to have confidence intervals that include zero than the others.
Both of these patterns are expected from the theory of change for this intervention \citep{AlsanMarcella2019DDMf}.

Although largely null, the results in Figure \ref{fig:fig2} for the White population are worth interrogating.
If the intervention is expected to be ineffective for White men, then the overall pattern of small but positive effects for the White population suggests that race is an effect modifier over and above any of the covariates in the profile.
Heuristically, we can use this in a sensitivity analysis for the other estimates by assuming that any unmeasured effect modification for the other target populations is no larger than that for the White population, in the spirit of negative controls \citep{LipsitchMarc2010NCAT}.
On the other hand, if the intervention can have effects for White men, these results suggest that Black doctors can be more effective than White doctors regardless of the race of their patient, which is an interesting hypothesis for further study.

%% file: onestep_sec7.tex
\section{Remarks}
\label{sec_remarks}

Methods for generalizing and transporting treatment effect estimates from a study sample to a target population typically involve weights.
In this paper, we provide a justification for estimating these weights in a single step and analyze the theoretical properties and empirical performance of the resulting one-step weights and estimators.

Weighting estimators with a multiplicative structure arise in numerous other areas of causal inference.
For example, estimators from the causal mediation, informative censoring, federated learning, longitudinal studies, and principal stratification literatures all traditionally involve weighting units by the products of multiple estimated weights.
Justifying the use of one-step weights for these settings and analyzing their properties are promising directions for future research.

%% file: onestep_supp.tex
\subsection{On the Identifying Assumptions for the Target Average Treatment Effect}

We now discuss the assumptions for identification of $\tau$, as given in Assumption 1 in Section 2.
We can weaken Assumption 1(b) to only require $E_{\mathbb{P}_1}\{Y(z)|\bm{X}\} = E_{\mathbb{P}_0}\{Y(z)|\bm{X}\}$, $z \in \{0,1\}$, i.e., conditional mean exchangeability of the potential outcomes across treatment groups. Also, we can weaken Assumption 1(d) to require $E_{\mathbb{T}}\{Y(1) - Y(0)|\bm{X}\} = E_{\mathbb{P}}\{Y(1)-Y(0)|\bm{X}\}$, i.e., conditional mean exchangeability of the treatment effect across the study and target populations. In this paper, we consider somewhat stronger assumptions as they allow us to identify general features of the distributions of the potential outcomes in the target population, e.g., $E_{\mathbb{T}}[g\{Y(z)\}]$, $z \in \{0,1\}$, and $\tilde{g}[E_{\mathbb{T}}\{Y(1)\},E_{\mathbb{T}}\{Y(0)\}]$ for measurable functions $g(\cdot)$ and $\tilde{g}(\cdot)$.

\textcolor{black}{By Assumption 1(d), controlling for the observed covariates is sufficient to remove differences in potential outcome distributions between the study and the target populations. Importantly, because Assumptions 1(c) and 1(d) are based completely on the distributions of $\{Y(1),Y(0),\bm{X}\}$ in the study and the target populations, they can pertain to a variety of generalization and transportation settings.
These include those where the study and target populations are disjoint (e.g., transportation) and those where the study population is nested within the target (e.g., generalization).}
\subsection{Connection to inverse propensity weighting}
\label{sec_connection}
\subsubsection{The dual optimization problem}
In this section, we formally discuss the equivalence between one-step balancing weights and inverse propensity modeling weights.
Recall that the primal optimization problem for the one-step balancing weights is given by
\begin{align}\label{minimal_app}
\argmin_{\bm{w}} \left\{  \sum_{i: Z_i=z}\psi(w_i) : \left|\sum_{i: Z_i=z} w_i B_k(\bm{X}_i) - \bar{B}^*_k    \right| \leq \delta_k, \; k = 1, 2, ..., K; \sum_{i:Z_i=z}w_i = 1
\right\}
\end{align}
\textcolor{black}{Throughout this section, we consider generalization in the nested study setting, because traditional inverse propensity weighting methods have only been developed for settings with a formal study selection process and selection indicator variable. 
Here, the study population $\mathcal{P}$ is nested within the target population $\mathcal{T}$, with $D$ indicating selection from $\mathcal{T}$ into $\mathcal{P}$.} 
We assume that the study sample of size $n$ is nested within a random sample of size $n^*$ from $\mathcal{T}$.
For a unit in group $Z=z$ with covariate vector $\bm{x}$, let   $\widehat{w}(\bm{x})$ and $w^{\text{IP}}(\bm{x})$ be its one-step balancing weight and the true inverse propensity weight, respectively, where the one-step balancing weights are obtained via (\ref{minimal_app}). 
In Theorem \ref{thm_dual1}, we show that the one-step balancing weights estimate the inverse propensity weights under a specific functional form of the inverse propensity weights and a loss function. 
Theorem \ref{thm_dual1} generalizes the results of \cite{wang2020minimal} and \cite{chattopadhyay2020balancing}, where similar connections between minimal and inverse propensity weights have been established in missing data and causal inference settings, respectively.
\begin{theorem} \normalfont
\label{thm_dual1}
\begin{enumerate}[label=(\alph*)]
\item[]
\item The dual problem of (\ref{minimal_app}) is equivalent to the empirical loss minimization problem with $L_1$ regularization:
\begin{align} \label{dual1}
& \underset{\bm{\lambda}}{\text{minimize}}
& & (n^*)^{-1}\sum_{i=1}^{n^*}\Big[-n^*\mathbbm{1}(Z_i=z)D_i\rho\{\bm{B}(\bm{X}_i)^\top \bm{\lambda} \} + \bm{B}(\bm{X}_i)^\top \bm{\lambda} \Big] + |\bm{\lambda}|^\top\bm{\delta},
\end{align}
where $\bm{\lambda}$ is a $K \times 1$ vector of dual variables corresponding to the $K$ balancing constraints, $|\bm{\lambda}|$ is the vector of component-wise absolute values of $\bm{\lambda}$, and $\rho(t) = t/n^* -  t(h')^{-1}(t) - h((h')^{-1}(t))$, with $h(t) = \psi(1/n^*-t)$.
\item If $\hat{\bm{w}}$ and $\bm{\lambda}^\dagger_z$ are solutions to the primal and dual forms of (\ref{minimal_app}), respectively, then for $i:Z_i=z$,
\begin{equation}
\widehat{w}_i = \rho'\{\bm{B}(\bm{X}_i)^\top \bm{\lambda}^\dagger_z\}.
\label{primaldual}
\end{equation}
\item If $\tilde{\bm{\lambda}}_z \in \argmin\limits_{\lambda} E_{\mathbb{T}}[-n^*\mathbbm{1}(Z_i=z)D_i\rho\{\bm{B}(\bm{X}_i)^\top \bm{\lambda} \} + \bm{B}(\bm{X}_i)^\top \bm{\lambda}|\bm{X}_i=\bm{x}]$, then $\tilde{\bm{\lambda}}_z$ satisfies
\begin{equation}
\rho'\{\bm{B}(\bm{x})^\top \tilde{\bm{\lambda}}_z\}= \left\{n^* \pi(\bm{x})\P_{\mathbb{P}}(Z_i=z|\bm{X}_i = \bm{x})\right\}^{-1}= (n^*)^{-1}w^{\text{IP}}(\bm{x}).
\end{equation}
\end{enumerate}
\end{theorem}
%
Parts (a) and (b) of Theorem \ref{thm_dual1} provide an alternative approach to obtain the one-step balancing weights via the dual form of the optimization problem in (\ref{minimal_app}), and part (c) links the form of the dual objective to inverse propensity weighting. Formally, suppose the true inverse propensity weights in group $Z=z$ have the functional form $w^{\text{IP}}(\bm{x}) = n^* \rho'\{\bm{B}(\bm{x})^\top \bm{\lambda}\}$ for some parameter $\bm{\lambda}$, and we estimate $\bm{\lambda}$ by minimizing the regularized empirical loss function in (\ref{dual1}). Equation \ref{primaldual} implies that the resulting estimated inverse propensity weights are proportional to the one-step balancing weights in the sense that $\widehat{w}^{\text{IP}}_i = n^* \widehat{w}_i$. 
Importantly, estimation of the inverse propensity weights here does not involve separate specifications of the treatment assignment and the study selection models.
Theorem \ref{thm_dual1} thus shows that our proposed method is equivalent to a one-step estimation method of the inverse propensity weights via a loss function that directly addresses covariate balance relative to the target and simultaneously ensures that the weights are less dispersed. 
\subsubsection{Proof of Theorem \ref{thm_dual1}}

We first consider the optimization problem without the normalization constraint and with $\delta_k>0$ for $k\in \{1,2,...,K\}$. The $k$th balancing constraint in the optimization problem in (9) can be written as,
\begin{align}
&\Big|\sum_{i: Z_i=z} w_i B_k(\bm{X}_i) - \bar{B}^*_k    \Big| \leq \delta_k \nonumber \\
\implies & \Big|\sum_{i=1}^{n^*}\mathbbm{1}(Z_i=z)D_i w_i B_k(\bm{X}_i) - (n^*)^{-1}\sum_{i=1}^{n^*} B_k(\bm{X}_i) \Big| \leq \delta_k \nonumber \\
\implies & \Big|\sum_{i=1}^{n^*}\big\{(n^*)^{-1} - \mathbbm{1}(Z_i=z)D_iw_i \big\} B_k(\bm{X}_i)\Big| \leq \delta_k \nonumber \\
\implies & \Big|\sum_{i=1}^{n^*}\xi_i B_k(\bm{X}_i)\Big| \leq \delta_k 
\end{align}
where $\xi_i = 1/n^*- \mathbbm{1}(Z_i=z)D_iw_i $. Thus, for the units in group $Z_i=z$ of the sample, $w_i = ( 1/n^* - \xi_i)$.
For the objective function $\sum_{i:Z_i=z}\psi(w_i)$, we have
\begin{equation}
\sum_{i:Z_i=z}\psi(w_i) = \sum_{i=1}^{n^*}\mathbbm{1}(Z_i=z)D_i h(\xi_i),
\end{equation}
where $h(x) = \psi(1/n^* - x)$.
We let $\underline{\bm{A}}$ be a $K \times n$ matrix whose $(i,j)$th element is $B_i(\bm{X}_j)$; $\underline{\bm{Q}} = \left(\underline{\bm{A}}^\top, -\underline{\bm{A}}^\top \right)^\top$; and $\bm{d} = \left(\bm{\delta}, \bm{\delta}\right)^T$. We can write the primal problem as
\begin{equation}\label{app_primal2}
    \begin{aligned} 
& \underset{\bm{w}}{\text{minimize}}
& & \sum_{i=1}^{n^*}\mathbbm{1}(Z_i=z)D_i h(\xi_i) \\
& \text{subject to}
& & \underline{\bm{Q}}\bm{\xi} \leq \bm{d} \\
\end{aligned}
\end{equation}
This gives us a convex optimization problem in $\bm{\xi}$ with linear constraints. The Lagrange dual function of the primal problem in Equation \ref{app_primal2} is given by $ \inf\limits_{\bm{\xi}}\{\sum_{i=1}^{n^*}\mathbbm{1}(Z_i=z)D_i h(\xi_i) + \bm{\lambda}^\top \underline{\bm{Q}}\bm{\xi}\} - \bm{\lambda}^\top \bm{d}$ (see, e.g., \citealt{boyd2004convex} Chapter 5). Let $\bm{Q}_i$ be the $i$th column of $\underline{\bm{Q}}$. The dual objective function can be written as
\begin{align*}
&-\sup\limits_{\bm{\xi}}\{-\sum_{i=1}^{n^*}\mathbbm{1}(Z_i=z)D_i h(\xi_i) - \bm{\lambda}^\top \underline{\bm{Q}}\bm{\xi}\} - \bm{\lambda}^\top \bm{d}\\ &= -\sum_{i=1}^{n^*} \sup\limits_{\xi_i}\{-(\bm{Q}^\top_i\bm{\lambda})\xi_i - \mathbbm{1}(Z_i=z)D_i h(\xi_i) \}- \bm{\lambda}^\top \bm{d}\\
&= \sum_{i=1}^{n^*}\{-h^*_i(-\bm{Q}^T_i\bm{\lambda})\} - \bm{\lambda}^{\top}\bm{d},
\end{align*}
where $h_i^{*}(\cdot)$ is the convex conjugate of $\mathbbm{1}(Z_i=z)D_ih(\cdot)$. Thus, the dual problem is given by
\begin{equation}
    \begin{aligned} 
& \underset{\bm{\lambda}}{\text{maximize}}
& & \sum_{i=1}^{n^*}\{-h^*_i(-\bm{Q}^T_i\bm{\lambda})\} - \bm{\lambda}^{\top}\bm{d} \\
& \text{subject to}
& & \bm{\lambda} \geq \bm{0}, \\
\end{aligned}
\end{equation}
Since the last $K$ components of $\bm{Q}_i$ are reflected versions of the first $K$ components, by symmetry we can write the dual problem as,
\begin{equation}
    \begin{aligned} 
& \underset{\bm{\lambda}}{\text{maximize}}
& & \sum_{i=1}^{n^*}\{-h^*_i(\bm{Q}^T_i\bm{\lambda})\} - \bm{\lambda}^{\top}\bm{d} \\
& \text{subject to}
& & \bm{\lambda} \geq \bm{0}, \\
\end{aligned}
\end{equation}
Now,
\vspace{-1.8\topsep}
\begin{align}
    h^*_i(t) &= \sup_{\xi_i}\{t\xi_i - \mathbbm{1}(Z_i=z)D_ih(\xi_i)\} \nonumber\\
    &= \sup_{w_i}\big[ \big\{ 1/n^* - \mathbbm{1}(Z_i=z)D_iw_i \big\}t - \mathbbm{1}(Z_i=z)D_i h \Big\{ 1/n^*- \mathbbm{1}(Z_i=z)D_iw_i \Big\}  \big] \nonumber\\
    &= \sup_{w_i}\big[ \big\{ 1/n^*- \mathbbm{1}(Z_i=z)D_iw_i \big\}t - \mathbbm{1}(Z_i=z)D_i h( 1/n^*-w_i)  \big]\nonumber\\
    & = \big\{ 1/n^* - \mathbbm{1}(Z_i=z)D_i\widehat{w}_i(t) \big\}t - \mathbbm{1}(Z_i=z)D_i h(1/n^*-\widehat{w}_i(t)\}, 
\end{align}
where $\widehat{w}_i(t)$ satisfies

\begin{equation}
\partial/\partial w_i \Big[\big\{ 1/n^* - \mathbbm{1}(Z_i=z)D_iw_i \big\}t - \mathbbm{1}(Z_i=z)D_i h( 1/n^*-w_i) \Big]\Big|_{w_i = \widehat{w}_i(t)}= 0. 
\end{equation}

Solving for $w_i$, we get $\widehat{w}_i(t) =  1/n^*-(h^{'})^{-1}(t)$.
Therefore, the dual problem boils down to,
\begin{equation}\label{dual}
    \begin{aligned} 
& \underset{\bm{\lambda}}{\text{minimize}}
& & \sum_{i=1}^{n^*} \Big\{-\mathbbm{1}(Z_i=z)D_i\rho(\bm{Q}^\top_i\bm{\lambda}) + 1/n^*\bm{Q}^\top_i\bm{\lambda}\Big \} +   \bm{\lambda}^\top\bm{d}  \\
& \text{subject to}
& & \bm{\lambda} \geq \bm{0}, \\
\end{aligned}
\end{equation}
where $\rho(t) =  t/n^* -t(h')^{-1}(t) + h\{(h')^{-1}(t)\}$. Note that $\rho'(t) =  1/n^* - (h')^{-1}(t)$. Therefore, 
\begin{equation}
    \widehat{w}_i(t) = \rho'(t). 
    \label{eq_optweights}
\end{equation}
Following the proof structure of Theorem 1 in \cite{wang2020minimal}, we write $\bm{\lambda} = (\bm{\lambda}^\top_{+}, \bm{\lambda}^\top_{-})^\top$, where $\bm{\lambda}_{+}$ and $\bm{\lambda}_{-}$ are $K \times 1$ vectors. Denoting $\bm{A}_i$ as the $i$th column of $\underline{\bm{A}}$, we write the dual objective as,
\begin{align}
  g(\bm{\lambda}) =  \sum_{i=1}^{n^*} \Big\{-\mathbbm{1}(Z_i=z)D_i\rho\{\bm{A}^\top_i(\bm{\lambda}_{+}- \bm{\lambda}_{-})\} +(n^*)^{-1}\bm{A}^\top_i(\bm{\lambda}_{+} - \bm{\lambda}_{-})\Big \} +   (\bm{\lambda}^\top_{+} + \bm{\lambda}^\top_{-})\bm{\delta}. 
  \label{eq_loss}
\end{align}
Denote $\bm{\lambda}^\dagger = (\bm{\lambda}^{\dagger\top}_{+}, \bm{\lambda}^{\dagger\top}_{-})^\top$ as the dual solution. 
Let, if possible, the $j$th component of $\bm{\lambda}^{\dagger}_+$ and $\bm{\lambda}^{\dagger}_-$ be both strictly positive, for some $j \in \{1,2,...,K\}$. Define,
\begin{equation}
    \bm{\lambda}^{\dagger \dagger} = \big(\bm{\lambda}^{\dagger}_{+\top} - (0,0,...,\underbrace{\min(\lambda^{\dagger}_{+,j},\lambda^{\dagger}_{-,j})}_{\text{$j$th place}},0,...,0)^\top, \bm{\lambda}^{\dagger\top}_{-} - (0,0,...,\underbrace{\min(\lambda^{\dagger}_{+,j},\lambda^{\dagger}_{-,j})}_{\text{$j$th place}},0,...,0)^\top  \big) 
\end{equation}
Notice that $g(\bm{\lambda}^{\dagger \dagger}) = g(\bm{\lambda}^\dagger) - 2\delta_j\min(\lambda^{\dagger}_{+,j},\lambda^{\dagger}_{-,j})< g(\bm{\lambda}^\dagger)$ since $\delta_j>0$. This leads to a contradiction since $\bm{\lambda}^{\dagger}$ minimizes $g(\bm{\lambda})$. This implies that at least one of $\lambda^{\dagger}_{+,j}$ $\lambda^{\dagger}_{-,j}$ equals zero. From Equation \ref{eq_loss} we see that the dual problem has the following unconstrained form.
\begin{equation}\label{dual_2}
\begin{aligned} 
& \underset{\bm{\lambda}}{\text{minimize}}
& & (n^*)^{-1} \sum_{i=1}^{n^*}\Big[-n^*\mathbbm{1}(Z_i=z)D_i\rho\{\bm{B}(\bm{X}_i)^\top \bm{\lambda} \} + \{ \bm{B}(\bm{X}_i)^\top \bm{\lambda} \}  \Big] + |\bm{\lambda}|^\top\bm{\delta},
\end{aligned}
\end{equation}
where $|\bm{\lambda}|$ is the vector of co-ordinate wise absolute values of $\bm{\lambda}$. 

Now, consider the primal problem 
\begin{equation}\label{minimal_exact}
\begin{aligned} 
& \underset{\bm{w}}{\text{minimize}}
& & \sum_{i: Z_i=z}\psi(w_i),\\
& \text{subject to}
& & \Big |\sum_{i: Z_i=z} w_i B_k(\bm{X}_i) - \bar{B}^*_k    \Big| \leq \delta_k, \; k=1,2,...,K_0.\\
& & & \Big |\sum_{i: Z_i=z} w_i B_k(\bm{X}_i) - \bar{B}^*_k    \Big| = 0, \; k=K_0+1,...,K.
\end{aligned} 
\end{equation}
Let $\tilde{\bm{B}}(\bm{x}) = (B_1(\bm{x}),...,B_{K_0}(\bm{x}))^\top$,  $\tilde{\tilde{\bm{B}}}(\bm{x}) = (B_{K_0+1}(\bm{x}),...,B_{K}(\bm{x}))^\top$, and $\tilde{\bm{\delta}} = (\delta_1,...,\delta_{K_0})^\top$. We note that for $B_k(\bm{x}) = 1$, the corresponding equality constraint boils down to the normalization constraint. Using similar steps as before, we see that the dual of (\ref{minimal_exact}) is
\begin{equation}\label{dual_exact}
\begin{aligned} 
& \underset{\tilde{\bm{\lambda}}, \bm{\nu}}{\text{minimize}}
& & (n^*)^{-1} \sum_{i=1}^{n^*}\Big[-n^*\mathbbm{1}(Z_i=z)D_i\rho\{\tilde{\bm{B}}(\bm{X}_i)^\top \tilde{\bm{\lambda}} + \tilde{\tilde{\bm{B}}}(\bm{X}_i)^\top \bm{\nu} \} + \{ \tilde{\bm{B}}(\bm{X}_i)^\top \tilde{\bm{\lambda}} + \tilde{\tilde{\bm{B}}}(\bm{X}_i)^\top \bm{\nu} \}  \Big] + |\tilde{\bm{\lambda}}|^\top\tilde{\bm{\delta}}.
\end{aligned}
\end{equation}
Let $\bm{\lambda} = (\tilde{\bm{\lambda}}^\top, \bm{\nu}^\top)^\top$. Since $\bm{B}(\bm{x}) = (\tilde{\bm{B}}(\bm{x})^\top, \tilde{\tilde{\bm{B}}}(\bm{x})^\top)^\top$ and  $\bm{\delta} = (\tilde{\bm{\delta}}^\top, \bm{0}^\top)^\top$, the dual problem in (\ref{dual_exact}) can be written as:
\begin{equation}\label{dual_3}
\begin{aligned} 
& \underset{\bm{\lambda}}{\text{minimize}}
& & (n^*)^{-1} \sum_{i=1}^{n^*}\Big[-n^*\mathbbm{1}(Z_i=z)D_i\rho\{\bm{B}(\bm{X}_i)^\top \bm{\lambda} \} + \{ \bm{B}(\bm{X}_i)^\top \bm{\lambda} \}  \Big] + |\bm{\lambda}|^\top\bm{\delta},
\end{aligned}
\end{equation}
which has the same form as in (\ref{dual_2}). This proves part (a) of Theorem 4.1.
Moreover, Equation \ref{eq_optweights} implies that the optimal solutions of the primal problem satisfies
\begin{equation}
\widehat{w}_i = \rho'\{\bm{B}(\bm{X}_i)^\top \bm{\lambda}^\dagger\},
\end{equation}
proving part (b). Finally, for part (c), we consider the conditional expected loss in (\ref{dual_2}).
\vspace{-1.8\topsep}
\begin{align}
&E_{\mathbb{T}}\Big[ -n^*\mathbbm{1}(Z_i=z)D_i\rho\{\bm{B}(\bm{X}_i)^\top \bm{\lambda} \} + \{ \bm{B}(\bm{X}_i)^\top \bm{\lambda} \} \Big|\bm{X}_i  \Big] \nonumber\\
&= E_{\mathbb{T}}\Big[-n^*\mathbbm{1}(Z_i=z)D_i\rho\{\bm{B}(\bm{X}_i)^\top \bm{\lambda} \}\Big|\bm{X}_i\Big] + \{ \bm{B}(\bm{X}_i)^\top \bm{\lambda} \} \nonumber\\
&= -n^*\rho\{\bm{B}(\bm{X}_i)^\top \bm{\lambda} \}\P_{\mathbb{T}}(Z_i=z|\bm{X}_i)\pi(\bm{X}_i) +  \{ \bm{B}(\bm{X}_i)^\top \bm{\lambda} \} =: \phi(\bm{\lambda})
\end{align}
We now minimize this expected loss $\phi(\bm{\lambda})$ wrt $\bm{\lambda}$. The minimizer $\bm{\lambda}^{*}$ satisfies the following:
\begin{equation}
\rho'\{\bm{B}(\bm{X}_i)^\top \bm{\lambda}^{*}\}= \left\{n^* \pi(\bm{X}_i)\P_{\mathbb{T}}(Z_i=z|\bm{X}_i)\right\}^{-1}
\end{equation}
This implies the solutions $\widehat{w}_i$ implicitly estimate the inverse propensity weights. This completes the proof.


\subsection{Connection to inverse odds weighting}

In this section, we consider the generalization problem in a nested design setting, where the target population is the population of study non-participants. As before, we denote $\mathcal{P}$ and $\mathcal{T}$ as the study and target population, respectively. Let $\mathcal{Q}$ be the population represented by the study participants and non-participants, with associated probability measure $\mathbb{Q}$ (and density $q$). Also, let $n^{**}$ be the total number of study participants and non-participants.
As before, $D$ is the indicator of being a study participant.
It follows that $\mathbb{P} = \mathbb{Q}|D=1$ and $\mathbb{T} = \mathbb{Q}|D=0$. In particular, we can write
\begin{equation}
\tau = E_{\mathbb{T}}\{Y(1) - Y(0)\} = E_{\mathbb{Q}}\{Y(1) - Y(0)|D=0\}
\end{equation}
We now derive a connection between the one-step balancing weights and the inverse odds weights in this setting. The $k$th balancing constraint in the optimization problem in (9) can be written as,
\begin{align}
&\Big|\sum_{i: Z_i=z} w_i B_k(\bm{X}_i) - \bar{B}^*_k    \Big| \leq \delta_k \nonumber \\
\implies & \Big|\sum_{i=1}^{n^{**}}\mathbbm{1}(Z_i=z)D_i w_i B_k(\bm{X}_i) - (n^*)^{-1}\sum_{i=1}^{n^{**}} (1-D_i)B_k(\bm{X}_i) \Big| \leq \delta_k \nonumber \\
\implies & \Big|\sum_{i=1}^{n^{**}}\big\{(n^*)^{-1} (1-D_i) - \mathbbm{1}(Z_i=z)D_iw_i \big\} B_k(\bm{X}_i)\Big| \leq \delta_k \nonumber \\
\implies & \Big|\sum_{i=1}^{n^*}\xi_i B_k(\bm{X}_i)\Big| \leq \delta_k 
\end{align}
where $\xi_i =(n^*)^{-1}(1-D_i) - \mathbbm{1}(Z_i=z)D_iw_i $. So, for the units in the group $Z_i=z$ of the sample, $w_i = \{(1-D_i)/n^* - \xi_i\} = -\xi_i$. 
For the objective function $\sum_{i:Z_i=z}\psi(w_i)$ (where $\psi$ is a convex function of the weights), we have
\begin{equation}
\sum_{i:Z_i=z}\psi(w_i) = \sum_{i=1}^{n^{**}}\mathbbm{1}(Z_i=z)D_i h(\xi_i) ,
\end{equation}
where $h(x) = \psi(-x)$. Now, let $\underline{\bm{A}}$ be a $K \times n$ matrix whose $(i,j)$th element is $B_i(\bm{X}_j)$; $\underline{\bm{Q}} = \left(\bm{A}^\top, -\bm{A}^\top \right)^T$; and $\bm{d} = \left(\bm{\delta}, \bm{\delta}\right)^T$. We can write the primal problem as
\begin{equation}\label{primal2}
    \begin{aligned} 
& \underset{\bm{w}}{\text{minimize}}
& & \sum_{i=1}^{n^{**}}\mathbbm{1}(Z_i=z)D_i h(\xi_i) \\
& \text{subject to}
& & \underline{\bm{Q}}\bm{\xi} \leq \bm{d} \\
\end{aligned}
\end{equation}
 gives us a convex optimization problem in $\bm{\xi}$ with linear constraints. 
Let $\bm{Q}_i$ be the $i$th column of $\underline{\bm{Q}}$. The corresponding dual problem is given by,
\begin{equation}
    \begin{aligned} 
& \underset{\bm{\lambda}}{\text{maximize}}
& & \sum_{i=1}^{n^{**}}\{-h^*_i(\bm{Q}^T_i\bm{\lambda})\} - \bm{\lambda}^{T}\bm{d} \\
& \text{subject to}
& & \bm{\lambda} \geq \bm{0} \\
\end{aligned}
\end{equation}
where,
\begin{align}
    h^*_i(t) &= \sup_{\xi_i}\{t\xi_i - \mathbbm{1}(Z_i=z)D_ih(\xi_i)\} \nonumber\\
    &= \sup_{w_i}\big[ \big\{(1-D_i)/n^* - \mathbbm{1}(Z_i=z)D_iw_i \big\}t - \mathbbm{1}(Z_i=z)D_i h \Big((1-D_i)/n^*- \mathbbm{1}(Z_i=z)D_iw_i \Big)  \big] \nonumber\\
    &= \sup_{w_i}\big[ \big\{(1-D_i)/n^* - \mathbbm{1}(Z_i=z)D_iw_i \big\}t - \mathbbm{1}(Z_i=z)D_i h(-w_i)  \big]\nonumber\\
    & = \big\{(1-D_i)/n^*- \mathbbm{1}(Z_i=z)D_i\widehat{w}_i(t) \big\}t - \mathbbm{1}(Z_i=z)D_i h(-\widehat{w}_i(t)) 
\end{align}
where $\widehat{w}_i(t)$ satisfies 
$$\partial/\partial w_i\Big[\big\{(1-D_i)/n^* - \mathbbm{1}(Z_i=z)D_iw_i \big\}t - \mathbbm{1}(Z_i=z)D_i h(-w_i) \Big]\Big|_{w_i = \widehat{w}_i(t)}= 0.$$ Solving for $w_i$, we get $\widehat{w}_i(t) = -(h^{'})^{-1}(t)$.\\
 
Therefore, the dual problem boils down to:
\begin{equation}\label{dual}
    \begin{aligned} 
& \underset{\bm{\lambda}}{\text{minimize}}
& & \sum_{i=1}^{n^{**}} \Big\{-\mathbbm{1}(Z_i=z)D_i\rho(\bm{Q}^\top_i\bm{\lambda}) +(n^*)^{-1}(1-D_i)\bm{Q}^\top_i\bm{\lambda}\Big \} +   \bm{\lambda}^\top\bm{d}  \\
& \text{subject to}
& & \bm{\lambda} \geq \bm{0} \\
\end{aligned}
\end{equation}

where $\rho(t) = -t(h')^{-1}(t) + h((h')^{-1}(t))$. Note that $\rho'(t) = -(h')^{-1}(t)$. Therefore,
\begin{equation}
    \widehat{w}_i(t) = \rho'(t).
\end{equation}
Using similar steps as in the proof of Theorem 4.1 we obtain the following form of the dual problem.
\begin{equation}\label{dual2.2}
\begin{aligned} 
& \underset{\bm{\lambda}}{\text{minimize}}
& &(n^{**})^{-1}\sum_{i=1}^{n^{**}}\Big[-n^{**}\mathbbm{1}(Z_i=z)D_i\rho\{\bm{B}(\bm{X}_i)^\top \bm{\lambda} \} + (n^*)^{-1} n^{**}(1-D_i)\{ \bm{B}(\bm{X}_i)^\top \bm{\lambda} \}  \Big] + |\bm{\lambda}|^\top\bm{\delta}
\end{aligned}
\end{equation}
Also, if $\widehat{w}_i$s are the optimal solutions of the primal problem and $\bm{\lambda}^\dagger$ is the optimal solution of the dual problem, then we have
\begin{equation}
\widehat{w}_i = \rho'\{\bm{B}(\bm{X}_i)^\top \bm{\lambda}^\dagger\}. 
\end{equation}
Let us consider the conditional expected loss in (\ref{dual2.2}).
\begin{align}
&E_{\mathbb{Q}}\Big[ -n^{**}\mathbbm{1}(Z_i=z)D_i\rho\{\bm{B}(\bm{X}_i)^\top \bm{\lambda} \} + (n^*)^{-1} n^{**}(1-D_i)\{ \bm{B}(\bm{X}_i)^\top \bm{\lambda} \} \Big|\bm{X}_i  \Big] \nonumber\\
&= E_{\mathbb{Q}}\Big[ -n^{**}\mathbbm{1}(Z_i=z)D_i\rho\{\bm{B}(\bm{X}_i)^\top \bm{\lambda} \}\Big|\bm{X}_i\Big] + (n^*)^{-1} n^{**} (1-\pi(\bm{X}_i)) \{ \bm{B}(\bm{X}_i)^\top \bm{\lambda} \} \nonumber\\
&= -n^{**}\rho\{\bm{B}(\bm{X}_i)^\top \bm{\lambda} \}\P_{\mathbb{Q}}(Z_i=z|\bm{X}_i)\pi(\bm{X}_i) + (n^*)^{-1} n^{**} \{1-\pi(\bm{X}_i)\} \{ \bm{B}(\bm{X}_i)^\top \bm{\lambda} \} \nonumber\\
 &=: \phi(\bm{\lambda})
\end{align}
We now minimize this expected loss $\phi(\lambda)$ wrt $\lambda$. The minimizer $\bm{\lambda}^{**}$ satisfies the following--
\begin{equation}
\rho'\{\bm{B}(\bm{X}_i)^\top \bm{\lambda}^{**}\}= \left\{1-\pi(\bm{X}_i)\right\}/\left\{n^* \pi(\bm{X}_i)\P_{\mathbb{Q}}(Z_i=z|\bm{X}_i)\right\}. 
\end{equation}
This implies, the solutions $\widehat{w}_i$ implicitly estimate the inverse odds weights.
{ \color{black}
\subsection{Connection to linear regression}
\label{sec_linearreg}

In this section, we discuss the connection between one-step weights and linear regression. For simplicity, we consider the generalization problem in a nested design setting, where the study sample is nested within a random sample of size $n^*$
 from the target population. Extensions of the results to other settings hold analogously. Also, without loss of generality, we focus on estimating $E_{\mathbb{T}}\{Y(1)\}$ (the results hold similarly for $E_{\mathbb{T}}\{Y(0)\}$).

Recall that, by Assumption \ref{assump1}, $m_1(\bm{x}) = E_{\mathbb{T}}\{Y(1)|\bm{X} = \bm{x}\} = E_{\mathbb{P}}\{Y(1)|\bm{X} = \bm{x}\}$. The linear regression imputation approach for generalization (our outcome modeling approach; see \cite{dahabreh2019generalizing}) first fits a linear outcome model $Y^{\text{obs}}_i = \beta_0 + \bm{\beta}^\top_1\bm{X}_i + \epsilon_i$ in the treatment group and estimates $m_1(\bm{x})$ as $\hat{m}_1(\bm{x}) = \hat{\beta}_0 + \hat{\bm{\beta}}^\top_1\bm{x}$, where the coefficients are estimated by ordinary least squares (OLS). The regression model can also incorporate other transformations of $\bm{X}$.
The regression imputation estimator of $E_{\mathbb{T}}\{Y(1)\}$ is given by,
\begin{equation}
    \hat{E}_{\mathbb{T}}\{Y(1)\} = \frac{1}{n^*} \sum_{i = 1}^{n^*}\{\hat{\beta}_0 + \hat{\bm{\beta}}^\top_1\bm{X}_i\} = \hat{\beta}_0 + \hat{\bm{\beta}}^\top_1\bar{\bm{X}}^*,
\end{equation}
where $\bar{\bm{X}}^*$ is the target profile, i.e., the mean of the covariates in the full sample.

We note that procedurally, this approach is equivalent to the multi-regression imputation (MRI) approach in \cite{chattopadhyay2023implied}. Hence, by Proposition 2 of \cite{chattopadhyay2023implied}, it follows that, $\hat{\beta}_0 + \hat{\bm{\beta}}^\top_1\bar{\bm{X}}^* = \sum_{i:Z_i=1}w_i Y^{\text{obs}}_i$, where the weights $w_i$ sum to one in the treatment group. 

Moreover, by Proposition 3 of the same paper, the weights $w_i$ are the weights of minimum variance that add up to one and \textit{exactly} balances the mean of the covariates in the treatment group, relative to the target profile $\bar{\bm{X}}^*$. In other words, these weights are equivalent to the one-step weights in \eqref{minimal}, where --- (i) the $L_2$ norm of the weights are minimized, (ii) The basis functions are identity functions, (iii) $\delta_k = 0$ for all $k$, and (iv) weights are allowed to be negative.

}
\subsection{Convergence of one-step weights}
\label{sec_appendix_convergence}

The equivalence in Theorem \ref{thm_dual1} allows us to establish several asymptotic properties of the one-step balancing weights and their resulting estimators. In this section, we formally show that under Assumption \ref{assump2} and  \ref{assump4}, the one-step weights converge uniformly to the inverse propensity weights.

For convenience, we restate the regularity conditions in Assumption 4 below.

\textbf{Assumption 4.} For $z \in \{0,1\}$, 
\begin{enumerate}[label=(\alph*)]

\item There exist constants $c_0$, $c_1$, $c_2$ with $0<c_0<1/2$ and $c_1<c_2<0$, such that $c_1 \leq n^*\rho{''}(v) \leq c_2$ for all $v$ in a neighborhood of $\bm{B}(\bm{x})^\top \bm{\lambda}^*_{1z}$. Also, $c_0 \leq 1/(n^*\rho'(v)) \leq 1-c_0$ for all $v = \bm{B}(\bm{x})^\top \bm{\lambda}$, $\bm{x} \in \mathcal{X}$, $\bm{\lambda}$. 
    
    \item $\sup\limits_{x \in \mathcal{X}}||\bm{B}(\bm{x})||_2 \leq CK^{1/2}$ and $||E_{\mathbb{T}}\{\bm{B}(\bm{X}) \bm{B}(\bm{X})^\top\}||_F \leq C$, for some constant $C>0$, where $||\cdot||_F$ denotes the Frobenius norm.
  
    \item $K = O\{(n^*)^\alpha\}$ for some $0<\alpha<2/3$.
    
    \item For some constant $C>0$, $\lambda_{\text{min}}\Big[ E_{\mathbb{T}}\{D\mathbbm{1}(Z = z)\bm{B}(\bm{X})\bm{B}(\bm{X})^\top\}\Big]>C$, where $\lambda_{\min}(\bm{A})$ denotes the smallest eigenvalue of $\bm{A}$.
    
    \item $||\bm{\delta}||_2 = O_P\left[K^{1/4}\left\{(\log K) / n^*\right\}^{1/2} + K^{- r_z + 1/2}\right]$.
\end{enumerate}

Assumption \ref{assump4}(a) bounds the slope and curvature of the function $\rho(\cdot)$, allowing us to translate the convergence of the dual solution to the convergence of the weights.
This condition is satisfied for typical choices of convex objective functions $\psi(\cdot)$, e.g., those corresponding to entropy balancing \citep{hainmueller2012balancing} and stable balancing weights \citep{zubizarreta2015stable}.
Assumption \ref{assump4}(b) restricts the rate of growth of the norm of the basis functions, and \ref{assump4}(c) specifies the rate of growth of the number of basis functions $K$ relative to $n^*$. Assumption \ref{assump4}(d) is a technical condition required to ensure non-singularity of the covariance matrices of the basis functions within each treatment group. Finally, \ref{assump4}(e) controls the degree of approximate balancing in terms of $K$ and $n^*$.

\begin{theorem}[Uniform convergence]\normalfont
\label{thm_conv1} Under Assumptions 1, 2, and \ref{assump4}, the one-step balancing weights in group $Z=z \in \{0,1\}$ satisfy
\begin{equation}
\sup\limits_{\bm{x} \in \mathcal{X}}|n^* \widehat{w}(\bm{x}) -w^{\text{IP}}(\bm{x})|  = O_P\left[K^{3/4}\left\{\left(\log K\right)/n^*\right\}^{1/2}+ K^{1 - r_z} \right] = o_P(1).    
\end{equation}
\end{theorem}
Below we provide a proof of Theorem \ref{thm_conv1}.
All probabilities and expectations in this proof are computed with respect to the probability measure $\mathbb{T}$. We focus on the primal problem in group $z$, for $z \in \{0,1\}$. For simplicity, let us denote $\tilde{Z} = \mathbbm{1}(Z_i=z)D_i$. The dual optimization problem can be written as,
\begin{equation}
\begin{aligned} 
& \underset{\bm{\lambda}}{\text{minimize}} \quad G(\bm{\lambda}), \text{ where}\\
&  G(\bm{\lambda})= (n^*)^{-1}\sum_{i=1}^{n^*}\Big[-n^*\tilde{Z}_i\rho\{\bm{B}(\bm{X}_i)^\top \bm{\lambda} \} + \{ \bm{B}(\bm{X}_i)^\top \bm{\lambda} \}  \Big] + |\bm{\lambda}|^\top\bm{\delta}.
\end{aligned}
\label{dual_cond}
\end{equation}
Let $\bm{\lambda}^\dagger$ be a solution to the dual problem. We first consider the $L_\infty$ distance between the scaled one-step balancing weights and inverse probability weights. In the following, we use $C,C',C''$ as generic positive constants whose values may change from one step to the next. 
\begin{align}
    &\sup\limits_{\bm{x} \in \mathcal{X}} |n^*\widehat{w}(\bm{x}) - w^{\text{IP}}(\bm{x}) | \nonumber\\
    &= \sup\limits_{\bm{x} \in \mathcal{X}}|n^*\rho'\{\bm{B}(\bm{x})^\top \bm{\lambda}^{\dagger}\} - n^* \rho'\{g^*_z(\bm{x})\}| \nonumber\\
     &\leq \sup\limits_{\bm{x} \in \mathcal{X}}|n^*\rho'\{\bm{B}(\bm{x})^\top \bm{\lambda}^{\dagger}\} - n^* \rho'\{\bm{B}(\bm{x})^\top \bm{\lambda}^*_{1z}\}| + \sup\limits_{\bm{x} \in \mathcal{X}}|n^*\rho'\{\bm{B}(\bm{x})^\top \bm{\lambda}^*_{1z}\} - n^* \rho'\{g^*_z(\bm{x})\}| \nonumber\\
     &\leq C\sup\limits_{\bm{x} \in \mathcal{X}}|\bm{B}(\bm{x})^\top (\bm{\lambda}^{\dagger} - \bm{\lambda}^*_{1z})| + C \sup\limits_{\bm{x} \in \mathcal{X}}|\bm{B}(\bm{x})^\top \bm{\lambda}^*_{1z} - g^*_z(\bm{x})| \nonumber\\
     & \leq CK^{1/2}||\bm{\lambda}^\dagger - \bm{\lambda}^*_{1z}||_2 +  O(K^{-r_z}). 
     \label{eq:sup_metric}
\end{align}
The first equality is due to Assumption \ref{assump2}. The first inequality is due to the triangle inequality. The second inequality follows from applying using the mean value theorem and Assumption \ref{assump4}(a). The final inequality is due to the Cauchy-Schwarz inequality and Assumptions 2 and \ref{assump4}(b). 

Let us now consider the following Lemma. 
\begin{lemma} \normalfont
There exists a dual solution $\bm{\lambda}^{\dagger}$ such that $||\bm{\lambda}^{\dagger} - \bm{\lambda}^*_{1z}||_2 = O_p[K^{1/4} \{(\log K)/n^*\}^{1/2} + K^{ - r_z + 1/2}]$.
\label{eq:coefrate}
\end{lemma}
Given Lemma \ref{eq:coefrate}, Equation \ref{eq:sup_metric} completes the proof of Theorem 4.2. So, it suffices to prove Lemma \ref{eq:coefrate}. For this, we require the following lemmas.
\begin{lemma}[Bernstein's inequality for random matrices]
\normalfont
Let $\underline{\bm{W}}_1, \underline{\bm{W}}_2,...,\underline{\bm{W}}_{n^*}$ be $d_1 \times d_2$ independent random matrices with $E(\underline{\bm{W}}_j) = \bm{0}$ and $||\underline{\bm{W}}_j||_2 \leq R_{n^*}$ a.s., where $||\cdot||_2$ denotes the spectral norm. Let $\sigma^2_{n^*} := \max\Big\{ ||\sum_{j=1}^{{n^*}}E(\underline{\bm{W}}_j\underline{\bm{W}}^\top_j)||_2, ||\sum_{j=1}^{{n^*}}E(\underline{\bm{W}}^\top_j\underline{\bm{W}}_j)||_2   \Big\}$. Then, for all $t \geq 0$,
\begin{equation}
    \P\Big( \left|\left|\sum_{j=1}^{{n^*}} \underline{\bm{W}}_j \right|\right|_2 \geq t \Big) \leq (d_1 + d_2) \exp\left[(t^2/2)/\{\sigma_{n^*}^2 + (R_{n^*}t)/3\}\right].
\end{equation}
\label{thm_bern}
\end{lemma}
\textit{Proof of Lemma \ref{thm_bern}}. See \cite{tropp2015introduction}.
\begin{lemma}
\normalfont
$\Big|\Big| (n^*)^{-1}\sum_{j=1}^{n^*}\{1-\tilde{Z}_j w^{\text{IP}}(\bm{X}_j)\}\bm{B}(\bm{X}_j)\Big|\Big|_2 = O_P[K^{1/4} \{(\log K)/n^*\}^{1/2}]$.
\label{lemma_bern1}
\end{lemma}
\textit{Proof of Lemma \ref{lemma_bern1}}. We will use Lemma \ref{thm_bern} to prove this. Let us denote 
\begin{equation}
\underline{\bm{W}}_j = (n^*)^{-1}\{1-\tilde{Z}_j w^{\text{IP}}(\bm{X}_j)\}\bm{B}(\bm{X}_j),\quad \text{for $j \in \{1,2,...,n^*\}$}.
\end{equation}
First, by unconfoundedness, $E(\underline{\bm{W}}_j) = \bm{0}$. Second, we have
\begin{align}
||\underline{\bm{W}}_j||_2 &= (n^*)^{-1}|1-\tilde{Z}_j n^*\rho'\{g^*_z(\bm{X}_i)\}|\times ||\bm{B}(\bm{X}_j)||_2 \nonumber\\
& \leq  [1 + |n^*\rho'\{g^*_z(\bm{X}_j)\}|](n^*)^{-1}\sup\limits_{\bm{x}\in \mathcal{X}}||\bm{B}(\bm{x})||_2 \nonumber\\
& \leq CK^{1/2}(n^*)^{-1}[C' + O(K^{-r_z})C] \leq (C' K^{1/2})/n^*.
\label{eq_Rn}
\end{align}
Here the second inequality is obtained by applying Assumption \ref{assump4}(b) on the second term in the product and using the mean value theorem on $n^*\rho'\{g^*_z(\bm{X}_j)\}$ about $\bm{B}(\bm{X}_j)^\top\bm{\lambda}^*_{1z}$, followed by assumptions \ref{assump4}(a) and 2. Next, we consider
\begin{align}
    \left|\left|\sum_{j=1}^{n^*}E(\underline{\bm{W}}^\top_j\underline{\bm{W}}_j)\right|\right|_2 &= \sum_{j=1}^{n^{*}}E\Big[(n^*)^{-2}\{1-\tilde{Z}_j w^{\text{IP}}(\bm{X}_j)\}^2 \bm{B}(\bm{X}_j)^\top \bm{B}(\bm{X}_j)\Big] \nonumber\\
    &\leq C (n^*)^{-2}\sum_{j=1}^{n^*}E \Big\{\bm{B}(\bm{X}_j)^\top \bm{B}(\bm{X}_j)\Big\} \nonumber \\
    & = C (n^*)^{-2}\text{trace}\Big[E \Big\{\bm{B}(\bm{X}_1) \bm{B}(\bm{X}_1)^\top \Big\}\Big].
    \label{eq_sigman0}
\end{align}
Here first inequality follows from applying the mean value theorem along with assumptions \ref{assump4}(a) and 2, similar to the steps in Equation \ref{eq_Rn}. Now, 
let $\lambda_1,...\lambda_K$ be the eigenvalues of a non-negative definite matrix $\bm{A}$. By the Cauchy-Schwarz inequality, $\text{trace}(\bm{A}) \leq K^{1/2} \left(\lambda^2_1 + ... + \lambda^2_K\right)^{1/2} = K^{1/2}||\bm{A}||_F$. Thus, from Equation \ref{eq_sigman0}, we get
\begin{equation}
    \left|\left|\sum_{j=1}^{n^*}E(\underline{\bm{W}}^\top_j\underline{\bm{W}}_j)\right|\right|_2 \leq (C K^{1/2})/n^* \left|\left|E \Big[\bm{B}(\bm{X}_1) \bm{B}(\bm{X}_1)^\top \Big]\right|\right|_F \leq (C' K^{1/2})/n^*,
    \label{eq_sigman1}
\end{equation}
where the last inequality holds due to Assumption \ref{assump4}(b). Next, we consider
\begin{align}
 \left|\left|\sum_{j=1}^{n^*}E(\underline{\bm{W}}_j\underline{\bm{W}}^\top_j)\right|\right|_2 & \leq   \sum_{j=1}^{n^{*}}\Big|\Big|E\Big[(n^*)^{-2}\{1-\tilde{Z}_j w^{\text{IP}}(\bm{X}_j)\}^2 \bm{B}(\bm{X}_j)  \bm{B}(\bm{X}_j)^\top\Big]\Big|\Big|_2 \nonumber\\
 &\leq C (n^*)^{-2} \sum_{j=1}^{n^*}||E\{\bm{B}(\bm{X}_j)  \bm{B}(\bm{X}_j)^\top\}||_2 \nonumber\\
 &\leq C (n^*)^{-2}\sum_{j=1}^{n^*}||E\{\bm{B}(\bm{X}_j)  \bm{B}(\bm{X}_j)^\top\}||_F \leq C''/n^*
 \label{eq_sigman2}
\end{align}
Here the first inequality is due to the triangle inequality. The second inequality follows from upper bounding $\{1-\tilde{Z}_j w^{\text{IP}}(\bm{X}_j)\}^2$ as before and using monotonicity of spectral norms. The third inequality holds since spectral norm is dominated by the Frobenius norm. Finally, the fourth inequality holds due to Assumption \ref{assump4}(b)f. Therefore, from equations \ref{eq_sigman1} and \ref{eq_sigman2} we get,
\begin{equation}
    \sigma^2_{n^*} := \max\Big\{ \left|\left|\sum_{j=1}^{n^*}E\{\underline{\bm{W}}_j\underline{\bm{W}}^\top_j\}\right|\right |_2, \left| \left|\sum_{j=1}^{n^*}E\{\underline{\bm{W}}^\top_j\underline{\bm{W}}_j\}\right| \right|_2   \Big\} \leq (C K^{1/2})/n^*.
\end{equation}
Using Lemma \ref{thm_bern}, we get,
\begin{equation}
    \P\left( ||\sum_{j=1}^{n^*} \underline{\bm{W}}_j ||_2 \geq t \right) \leq (K + 1) \exp\left[(t^2/2)/\left\{C K^{1/2} (n^*)^{-1} + C' K^{1/2} t (3n^*)^{-1}\right\}\right]
    \label{eq:rate1}
\end{equation}
Finally, we observe that due to Assumption \ref{assump4}(c), the right hand side of Equation \ref{eq:rate1} goes to zero if $t = \bar{C} K^{1/4} (\log K)^{1/2} (n^*)^{-1/2}$, for some constant $\bar{C}>0$. This implies, $||\sum_{j=1}^{n^*} \underline{\bm{W}}_j ||_2 = O_P\left\{ K^{1/4} (\log K)^{1/2} (n^*)^{-1/2}\right\}$. This completes the proof.

\begin{lemma} \normalfont
With probability tending to one, $\lambda_{\min}\Big(\sum_{j:\tilde{Z}_j = 1} (n^*)^{-1}\bm{B}(\bm{X}_j)\bm{B}(\bm{X}_j)^\top  \Big)\geq \tilde{C}$ for some constant $\tilde{C}>0$.
\label{lemma_eigen}
\end{lemma}

\textit{Proof of Lemma \ref{lemma_eigen}}. Let $\bm{D} := \sum_{j:\tilde{Z}_j = 1}(n^*)^{-1} \bm{B}(\bm{X}_j)\bm{B}(\bm{X}_j)^\top = (n^*)^{-1}\sum_{j=1}^{n^*} \{\tilde{Z}_j \bm{B}(\bm{X}_j)\}\{\tilde{Z}_j \bm{B}(\bm{X}_j)^\top\}$ and
$\bm{D}^*:= E\{\tilde{Z}_j\bm{B}(\bm{X}_j)\bm{B}(\bm{X}_i)^\top\}$. We will first use Lemma \ref{thm_bern} to show that $||\bm{D} - \bm{D}^*||_2 = o_P(1)$.
To this end, denote $\underline{\bm{W}}_j = (n^*)^{-1}[\{\tilde{Z}_j \bm{B}(\bm{X}_j)\}\{\tilde{Z}_i \bm{B}(\bm{X}_j)^\top\} - E\{\tilde{Z}_j\bm{B}(\bm{X}_j)\bm{B}(\bm{X}_j)^\top\}]$. By construction, $E(\underline{\bm{W}}_j) = \bm{0}$. Moreover,
\begin{align}
    ||\{\tilde{Z}_j \bm{B}(\bm{X}_j)\}\{\tilde{Z}_j \bm{B}(\bm{X}_j)^\top\}||_2 &\leq   ||\{\tilde{Z}_j \bm{B}(\bm{X}_j)\}\{\tilde{Z}_j \bm{B}(\bm{X}_j)^\top\}||_F \nonumber\\
    & \leq C'\sup_{\bm{x}\in \mathcal{X}}||\bm{B}(\bm{x})||^2_2 \leq C''K,
\end{align}
where the last inequality holds due to Assumption \ref{assump4}(b). Also, 
\begin{align}
    ||E\{\tilde{Z}_j \bm{B}(\bm{X}_j)\bm{B}(\bm{X}_j)^\top\}||_2 &\leq ||E\{\bm{B}(\bm{X}_j)\bm{B}(\bm{X}_j)^\top\}||_2 \nonumber \\
    & \leq ||E\{\bm{B}(\bm{X}_j)\bm{B}(\bm{X}_j)^\top\}||_F \leq C',
\end{align}
where the first inequality is due to the monotonicity of the spectral norm and the last inequality holds due to Assumption \ref{assump4}(b). This implies,
\begin{align}
    ||\underline{\bm{W}}_j||_2 \leq \{C(K+1)\}/n^*,
\end{align}
for some constant $C>0$. Next, we compute $\sigma^2_{n^*}$. After some algebra, it follows that,
\begin{align}
    &||\sum_{j=1}^{n^*}E(\underline{\bm{W}}_j\underline{\bm{W}}^\top_j)||_2 \nonumber\\
    &\leq \sum_{j=1}^{n^*}||E(\underline{\bm{W}}_j\underline{\bm{W}}^\top_j)||_2 \nonumber\\
    &\leq(n^*)^{-1}\Big( CK||E\{ \bm{B}(\bm{X}_1)\bm{B}(\bm{X}_1)^\top\}||_2 + ||E\{\tilde{Z}_1\bm{B}(\bm{X}_1)\bm{B}(\bm{X}_1)^\top\}E\{\tilde{Z}_1\bm{B}(\bm{X}_1)\bm{B}(\bm{X}_1)^\top\}   ||_2\Big) \nonumber\\
    &\leq (n^*)^{-1}\Big[ CK||E\{ \bm{B}(\bm{X}_1)\bm{B}(\bm{X}_1)^\top\}||_2 + || E\{ \bm{B}(\bm{X}_1)\bm{B}(\bm{X}_1)^\top\} ||^2_2 \Big] \nonumber\\
    &\leq \{C'(K+1)\}/n^*,
\end{align}
for some large $C'>0$. Here the first inequality is due to the triangle inequality; the second inequality is due to Assumption \ref{assump4}(b) and monotonicity of the spectral norm; the third inequality is due to the submultiplicativity of the spectral norm; and the final inequality is due to Assumption \ref{assump4}(b). Now, since $\underline{\bm{W}}_j$ is symmetric we have
\begin{align}
    ||\sum_{j=1}^{n^*}E(\underline{\bm{W}}^\top_j\underline{\bm{W}}_j)||_2 \leq \{C'(K+1)\}/n^*,
\end{align}
implying $\sigma^2_{n^*} \leq \{C'(K+1)\}/n^*$. Therefore, by Lemma \ref{thm_bern}, we get
\begin{align}
    \P\left( \left|\left|\sum_{j=1}^{n^*} \underline{\bm{W}}_j \right|\right|_2 \geq t \right) &\leq 2K\exp \left[ (t^2/2)/\left\{C'(K+1)(n^*)^{-1} + C(K+1)t (2n^*)^{-1}\right\}\right]\notag\\
    &\leq 2K \exp\left[n^*t^2/\{C'' K(1+t)\}\right], \
\end{align}
for a large constant $C''>0$. We note that the right hand side goes to zero for $t = \bar{C} \{(K \log K) / n^*\}^{1/2}$ for a constant $\bar{C}>0$. Therefore, we have 
\begin{align}
    ||\bm{D} - \bm{D}^*||_2 = O_p \left[\left\{(K \log K)/n^*\right\}^{1/2}\right]= o_p(1),
\end{align}
where the last equality holds due to Assumption \ref{assump4}(c). Now, Weyl's inequality implies 
\begin{equation}
    \lambda_{\min}(\bm{D}) \geq \lambda_{\min}(\bm{D}^*) - ||\bm{D} - \bm{D}^*||_2 \geq C - ||\bm{D} - \bm{D}^*||_2 ,
\end{equation}
where the last inequality holds due to Assumption \ref{assump4}(d). Since $||\bm{D} - \bm{D}^*||_2 = o_p(1)$, we have for $n^*$ large enough,     $\lambda_{\min}(\bm{D}) \geq C/2 >0$. This completes the proof.

\textit{Proof of Lemma \ref{eq:coefrate}}. We follow the proof structure of \cite{fan2016improving} and \cite{wang2020minimal}. All the subsequent probabilities and expectations are taken with respect to $\mathbb{T}$. First, let $r = C^* \left\{ (K^{1/4} \log K) / n^* + K^{-r_z + 1/2}\right\}$ for a sufficiently large constant $C^*>0$. Let $\bm{\Delta} = \bm{\lambda} - \bm{\lambda}^*_{1z}$. Also, let $\mathcal{C} = \{\bm{\Delta}\in \mathbb{R}^K: ||\bm{\Delta}||_2 \leq r \}$. To show that there exists a $\bm{\lambda}^{\dagger}$ such that $||\bm{\lambda}^{\dagger} - \bm{\lambda}^*_{1z}||_2 = O_p\left\{ (K^{1/4} \log K) / n^* + K^{-r_z + 1/2}\right\}$, it is enough to show that there exists a $\bm{\Delta}^\dagger \in \mathbb{R}^K$ such that $\P(\bm{\Delta}^\dagger \in \mathcal{C}) \xrightarrow{n^* \to \infty} 1$ .

Now, the dual objective can be written as,
\begin{equation}
G(\bm{\lambda}^*_{1z} + \bm{\Delta})=(n^*)^{-1} \sum_{i=1}^{n^*}\Big[-n^*\tilde{Z}_i\rho\{\bm{B}(\bm{X}_i)^\top (\bm{\lambda}^*_{1z} + \bm{\Delta}) \} + \{ \bm{B}(\bm{X}_i)^\top (\bm{\lambda}^*_{1z} + \bm{\Delta}) \}  \Big] + |\bm{\lambda}^*_{1z} + \bm{\Delta}|^\top\bm{\delta} .   
\end{equation}
Since $f(\cdot)$ is convex, $\rho(\cdot)$ is concave. It follows that $G(\bm{\lambda}^*_{1z} + \bm{\Delta})$ is convex in $\bm{\Delta}$. Moreover, $G(\bm{\lambda}^*_{1z} + \bm{\Delta})$ is also continuous in $\bm{\Delta}$. Therefore, to show $\P(\bm{\Delta}^\dagger \in \mathcal{C}) \xrightarrow{n^* \to \infty} 1$, it is enough show that 
\begin{equation}
    \P\Big(\inf\limits_{\bm{\Delta} \in \partial\mathcal{C}}G(\bm{\lambda}^*_{1z} + \bm{\Delta}) - G(\bm{\lambda}^*_{1z})>0\Big) \xrightarrow{n^* \to \infty} 1,
    \label{eq_boundary}
\end{equation}
where $\partial\mathcal{C}$ is the boundary set of $\mathcal{C}$ given by $\partial\mathcal{C} =  \{\bm{\Delta}\in \mathbb{R}^K: ||\bm{\Delta}||_2 = r \}$.

Now, fix $\bm{\Delta} \in \partial\mathcal{C}$. Using multivariate Taylor's theorem, we seen that for some intermediate $\tilde{\bm{\lambda}}$,  
\begin{align}
    &G(\bm{\lambda}^*_{1z} + \bm{\Delta}) - G(\bm{\lambda}^*_{1z}) \nonumber\\
    &= \bm{\Delta}^\top \Big[ (n^*)^{-1}\sum_{i=1}^{n^*}\{-n^*\tilde{Z}_i \rho'\{\bm{B}(\bm{X}_i)^\top \bm{\lambda}^*_{1z}\}\bm{B}(\bm{X}_i) + \bm{B}(\bm{X}_i) \}\Big] \nonumber\\
    & + \bm{\Delta}^\top \Big[ (n^*)^{-1}\sum_{i=1}^{n^*}\{-n^*\tilde{Z}_i \rho''(\bm{B}(\bm{X}_i)^\top \tilde{\bm{\lambda}})\bm{B}(\bm{X}_i)\bm{B}(\bm{X}_i)^\top\} \Big]\bm{\Delta}/2 + (|\bm{\lambda}^*_{1z} + \bm{\Delta}| - |\bm{\lambda}^*_{1z}|)^\top\bm{\delta} \nonumber\\
    &\geq -||\bm{\Delta}||_2 \Big|\Big|(n^*)^{-1}\sum_{i=1}^{n^*}\{-n^*\tilde{Z}_i \rho'\{\bm{B}(\bm{X}_i)^\top \bm{\lambda}^*_{1z}\}\bm{B}(\bm{X}_i) + \bm{B}(\bm{X}_i) \}\Big |\Big|_2 + (\bm{\Delta}^\top \underline{\bm{M}}\bm{\Delta})/2 - |\bm{\Delta}|^\top \bm{\delta},
\end{align}
where $\underline{\bm{M}} = (n^*)^{-1}\sum_{i=1}^{n^*}\{-n^*\tilde{Z}_i \rho''(\bm{B}(\bm{X}_i)^\top \tilde{\bm{\lambda}})\bm{B}(\bm{X}_i)\bm{B}(\bm{X}_i)^\top\}$. Here the last inequality is due to the Cauchy-Schwarz inequality (for the first term) and the triangle inequality (for the third term). By Cauchy Schwarz, we get,
\begin{align}
&  G(\bm{\lambda}^*_{1z} + \bm{\Delta}) - G(\bm{\lambda}^*_{1z})   \nonumber\\
& \geq (\bm{\Delta}^\top \underline{\bm{M}}\bm{\Delta})/2 -r\Big( \Big|\Big| =(n^*)^{-1}\sum_{i=1}^{n^*}\{-n^*\tilde{Z}_i \rho'\{\bm{B}(\bm{X}_i)^\top \bm{\lambda}^*_{1z}\}\bm{B}(\bm{X}_i) + \bm{B}(\bm{X}_i) \}\Big |\Big|_2 + ||\bm{\delta}||_2\Big),
\end{align}
since $||\bm{\Delta}||_2 = r$. Now, 
\begin{align}
    &\Big|\Big| (n^*)^{-1}\sum_{i=1}^{n^*}\{-n^*\tilde{Z}_i \rho'\{\bm{B}(\bm{X}_i)^\top \bm{\lambda}^*_{1z}\}\bm{B}(\bm{X}_i) + \bm{B}(\bm{X}_i) \}\Big |\Big|_2 \nonumber\\
    &\leq \Big|\Big| (n^*)^{-1}\sum_{i=1}^{n^*}\{-\tilde{Z}_i n^*\rho'\{g^*_z(\bm{X}_i)\}\bm{B}(\bm{X}_i) + \bm{B}(\bm{X}_i) \}\Big |\Big|_2 + \nonumber\\
    &\Big|\Big| (n^*)^{-1}\sum_{i=1}^{n^*}-\tilde{Z}_i [n^*\rho'\{g^*_z(\bm{X}_i)\} - n^*\rho'\{\bm{B}(\bm{X}_i)^\top \bm{\lambda}^*_{1z}\}]\bm{B}(\bm{X}_i) \Big |\Big|_2 \nonumber\\
    &\leq \Big|\Big|(n^*)^{-1}\sum_{i=1}^{n^*}\{1-\tilde{Z}_i w^{\text{IP}}(\bm{X}_i)\}\bm{B}(\bm{X}_i)\Big|\Big|_2 +  \sup\limits_{\bm{x}\in \mathcal{X}}|g^*_z(\bm{x}) - \bm{B}(\bm{x})^\top \bm{\lambda}^*_{1z}| \Big|\Big|(n^*)^{-1}\sum_{i=1}^{n^*}-\tilde{Z}_i\bm{B}(\bm{X}_i)\Big|\Big|_2 \nonumber\\
    & \leq \Big|\Big|(n^*)^{-1}\sum_{i=1}^{n^*}\{1-\tilde{Z}_i w^{\text{IP}}(\bm{X}_i)\}\bm{B}(\bm{X}_i)\Big|\Big|_2 +   O(K^{-r_z})(n^*)^{-1}||\bm{B}(\bm{X}_i)||_2 \nonumber\\
    &\leq O_P\left\{(K^{1/4} \log K)/(n^*)^{1/2}\right\} + C K^{- r_z + 1/2}=  O_P\left\{( K^{1/4} \log K)/(n^*)^{1/2}+ K^{- r_z + 1/2}\right\}.
    \label{eq_normorder}
\end{align}
Here, the first inequality is due to the triangle inequality, and the second inequality is due to the mean value theorem. The third inequality is due to Assumption \ref{assump2} and the triangle inequality. Finally, the fourth inequality is due to Lemma \ref{lemma_bern1} and Assumption \ref{assump4}(b). Equation \ref{eq_normorder} combined with Assumption \ref{assump4}(e) implies that with probability tending to one,
\begin{align}
& G(\bm{\lambda}^*_{1z} + \bm{\Delta}) - G(\bm{\lambda}^*_{1z}) \nonumber \\  
& \geq (\bm{\Delta}^\top \underline{\bm{M}}\bm{\Delta})/2 -r O_P\left\{(K^{1/4} \log K)/(n^*)^{1/2} + K^{-r_z + 1/2}\right\} \nonumber\\
& = (2n^*)^{-1}\sum_{i=1}^{n^*}[-n^*\tilde{Z}_i \rho''(\bm{B}(\bm{X}_i)^\top \tilde{\bm{\lambda}})\{\bm{\Delta}^\top\bm{B}(\bm{X}_i)\}^2] - r O_P\left\{(K^{1/4} \log K)/(n^*)^{1/2} + K^{-r_z + 1/2}\right\} \nonumber\\
& \geq C (n^*)^{-1}\sum_{i=1}^{n^*}\tilde{Z}_j\{\bm{\Delta}^\top\bm{B}(\bm{X}_i)\}^2 - r O_P\left\{(K^{1/4} \log K)/(n^*)^{1/2} + K^{-r_z + 1/2}\right\} \nonumber\\
& = C \bm{\Delta}^\top \Big\{\sum_{i:\tilde{Z}_i = 1} (n^*)^{-1}\bm{B}(\bm{X}_i)\bm{B}(\bm{X}_i)^\top  \Big\}\bm{\Delta} - r O_P\left\{(K^{1/4} \log K)/(n^*)^{1/2} + K^{-r_z + 1/2}\right\} \nonumber\\
&\geq C r^2 \lambda_{\min}\Big(\sum_{i:\tilde{Z}_i = 1} (n^*)^{-1}\bm{B}(\bm{X}_i)\bm{B}(\bm{X}_i)^\top  \Big) - r O_P\left\{(K^{1/4} \log K)/(n^*)^{1/2} + K^{-r_z + 1/2}\right\} \nonumber\\
&\geq C''r^2-rO_P\left\{(K^{1/4} \log K)/(n^*)^{1/2} + K^{-r_z + 1/2}\right\} \nonumber\\
& = C''(C^*)^2\Big(K^{1/4}\{(\log K)/n^*\}^{1/2} + K^{-r_z + 1/2}\Big)^2 \notag \\
&\quad \quad \quad - C^*O_P\Big(\Big\{K^{1/4}\{(\log K)/n^*\}^{1/2} + K^{-r_z + 1/2}\Big\}^2\Big)>0.
\end{align}
Here the second inequality holds due to Assumption \ref{assump4}(a). The third inequality holds since for a square matrix $\bm{A}$, $\bm{x}^\top \bm{A} \bm{x} \geq \lambda_{\min}(A) ||\bm{x}||^2_2$. The fourth inequality is due to Lemma \ref{lemma_eigen}. Finally, the fifth inequality holds for a choice of $C^*$ large enough. This completes the proof of Lemma \ref{eq:coefrate} and Theorem 4.2.

\subsection{Proofs of propositions and theorems}

\subsubsection{Proof of Theorem 1}

We first show that when Assumption \ref{assump3} holds, $\sum_{i:Z_i=z}\widehat{w}_i Y_i(z)  \xrightarrow[n \to \infty]{P}  E_{\mathbb{T}}\{Y(z)\}$, for $z \in \{0,1\}$. For simplicity, we denote $\tilde{Z}_i = D_i\mathbbm{1}(Z_i=z)$. Also, denote $\bar{\bm{B}}^* = (n^*)^{-1}\sum_{i=1}^{n^*}\bm{B}(\bm{X}_i)$.
Writing $Y_i(z) = m_z(\bm{X}_i) + \epsilon_{iz}$, where $E_{\mathbb{T}}(\epsilon_{iz}|\bm{X}_i) = 0$, we get the following decomposition.
\begin{align}
&\Big|\sum_{i:Z_i=z}\widehat{w}_i Y_i(z) - E_{\mathbb{T}}\{Y(z)\} \Big| \nonumber\\
&\leq \Big|\sum_{i=1}^{n^*}\tilde{Z}_i \widehat{w}_i\{m_z(\bm{X}_i) - \bm{\lambda}^{*\top}_{2z}\bm{B}(\bm{X}_i)\} \Big| + \Big| \bm{\lambda}^{*\top}_{2z}\Big\{ \sum_{i:Z_i=z}\widehat{w}_i \bm{B}(\bm{X}_i) - \bar{\bm{B}}^* \Big\} \Big| \nonumber\\
& + \Big| \bm{\lambda}^{*\top}_{2z} \bar{\bm{B}} - (n^*)^{-1}\sum_{i=1}^{n^*}m_z(\bm{X}_i) \Big| + \Big|(n^*)^{-1}\sum_{i=1}^{n^*}m_z(\bm{X}_i) - E_{\mathbb{T}}\{m_z(\bm{X}_i)\} \Big| + \Big|\sum_{i=1}^{n^*}\tilde{Z}_i\widehat{w}_i\epsilon_{iz} \Big|  \nonumber\\
&\leq \sup\limits_{\bm{x}\in \mathcal{X}}|m_z(\bm{x}) - \bm{\lambda}^{*\top}_{2z}\bm{B}(\bm{x})| \sum_{i=1}^{n^*}\tilde{Z}_i|\widehat{w}_i| + |\bm{\lambda}^*_{2z}|^\top\bm{\delta} + \sup\limits_{\bm{x}\in \mathcal{X}}|m_z(\bm{x}) - \bm{\lambda}^{*\top}_{2z}\bm{B}(\bm{X}_i)| \nonumber \\
& + o_P(1) + \Big|\sum_{i=1}^{n^*}\tilde{Z}_i\widehat{w}_i\epsilon_{iz} \Big|\nonumber\\
&\leq O(K^{-s_{z}}) |(n^*)^{-1}\sum_{i=1}^{n*}\tilde{Z}_in^*\rho'\{\bm{\lambda}^{\dagger\top}\bm{B}(\bm{X}_i)\}|+ ||\bm{\lambda}^*_{2z}||_2 ||\bm{\delta}||_2 + O(K^{-s_{z}}) + o_P(1)+ \Big|\sum_{i=1}^{n^*}\tilde{Z}_i\widehat{w}_i\epsilon_{iz} \Big|\nonumber\\
& = o_P(1) +\Big|(n^*)^{-1}\sum_{i=1}^{n^*}\tilde{Z}_in^*\rho'\{\bm{\lambda}^{\dagger\top}\bm{B}(\bm{X}_i)\}\epsilon_{iz}\Big|.
\end{align}
Here the first inequality is due to the triangle inequality. In the second inequality, we bound the imbalances $|\sum_{i:Z_i=z}w_i \bm{B}(\bm{X}_i) - \bar{\bm{B}}^*|$ by $\bm{\delta}$ (component-wise). The third inequality holds due to the Cauchy-Schwarz inequality (for the second term) and Assumption \ref{assump3} (for the first and third terms). The final equality is due to assumptions 3 and \ref{assump4}(a). Now,
\begin{align}
(1 - c_0)^{-1}(n^*)^{-1}\sum_{i=1}^{n^*}\tilde{Z}_i\epsilon_{iz} \leq     (n^*)^{-1}\sum_{i=1}^{n^*}\tilde{Z}_in^*\rho'\{\bm{\lambda}^{\dagger\top}\bm{B}(\bm{X}_i)\}\epsilon_{iz} \leq c_0^{-1}(n^*)^{-1}\sum_{i=1}^{n^*}\tilde{Z}_i\epsilon_{iz}.
\end{align}
Both the upper and and lower bounds converge to a constant times $E_{\mathbb{T}}(\tilde{Z}_i\epsilon_{iz}) = E_{\mathbb{T}}\{\pi(\bm{X}_i)\P_{\mathbb{P}}(Z_i = 1|\bm{X}_i)E_{\mathbb{T}}(\epsilon_{iz}|\bm{X}_i)\} =0$, by Assumption 1. Therefore, $\Big|(n^*)^{-1}\sum_{i=1}^{n^*}\tilde{Z}_in^*\rho'\{\bm{\lambda}^{\dagger\top}\bm{B}(\bm{X}_i)\}\epsilon_{iz}\Big| = o_P(1)$. 

We now show that when Assumption \ref{assump2} holds, $\sum_{i:Z_i=z}\widehat{w}_i Y_i(z)  \xrightarrow[n \to \infty]{P}  E_{\mathbb{T}}\{Y(z)\}$.
\begin{align}
&|\sum_{i:Z_i=z}w_i Y^{\mathrm{obs}}_i - E_{\mathbb{T}}\{Y(z)\}| \nonumber\\
&\leq \Big|\sum_{i:Z_i=z}w_i Y_i(z) - (n^*)^{-1}\sum_{i:Z_i=z}w^{\text{IP}}_iY_i(z)\Big| + \Big|(n^*)^{-1}\sum_{i:Z_i=z}w^{\text{IP}}_iY_i(z) - E_{\mathbb{T}}\{Y(z)\} \Big| \nonumber \\
& \leq  \sup\limits_{\bm{x} \in \mathcal{X}}\Big|n^* w(\bm{x}) - w^{\text{IP}}(\bm{x})\Big| \Big\{ (n^*)^{-1}\sum_{i:Z_i=z}|Y_i(z)| \Big\}+ o_P(1) \nonumber \\
& = o_P(1),
\end{align}
where the last step holds due to Theorem 4.2. This completes the proof.

\subsubsection{Proof of Theorem 2}
\label{sec_appendix_Normality}

For convenience, we restate Assumption \ref{assump5} below.

\textbf{Assumption 5.} For $z \in \{0,1\}$, 
\begin{enumerate}[label=(\alph*)]

\item $E_{\mathbb{T}}\{Y^2(z)\}<\infty$.

\item Let $g^*_z(\cdot) \in \mathcal{G}_z$. $\mathcal{G}_z$ satisfies $\log N_{[]}\{\epsilon, \mathcal{G}_z, L_2(P)\} \leq C_1(1/\epsilon)^{1/k_1}$ for some constants $C_1>0$ and $k_1>1/2$, where $N_{[]}\{\epsilon, \mathcal{G}_z, L_2(P)\}$ is the covering number of $\mathcal{G}_z$ by epsilon brackets.
    
\item Let $m_z(\cdot) \in \mathcal{M}_z$. $\mathcal{M}_z$ satisfies $\log N_{[]}\{\epsilon, \mathcal{M}_z, L_2(P)\} \leq C_2(1/\epsilon)^{1/k_2}$ for some constants $C_2>0$ and $k_2>1/2$, , where $N_{[]}\{\epsilon, \mathcal{M}_z, L_2(P)\}$ is the covering number of $\mathcal{M}_z$ by epsilon brackets.

\item $(n^{*})^{\{2(r_z + s_{z} - 0.5)\}^{-1}} = o(K)$, where $r_z, s_z$ are the constants in assumptions 2 and 3, respectively. 
    
\end{enumerate}

The conditions in Assumption \ref{assump5} are similar to Assumption \ref{assump2} in \cite{wang2020minimal} and Assumption 4.1 in \cite{fan2016improving}. In particular, Assumption \ref{assump5}(a) ensures existence of the second moment of $Y(z)$ with respect to the target distribution. Assumptions \ref{assump5}(b) and (c) control the complexity of the function classes $\mathcal{G}_z$ and $\mathcal{M}_z$. Finally, Assumption \ref{assump5}(d) puts further restriction on the growth rate of the number of basis functions $K$ as a function of $n^*$.

All probabilities and expectations in this proof are computed with respect to the probability measure $\mathbb{T}$. We first decompose the Hajek estimator $T - \tau = \{\sum_{i:Z_i = 1}\widehat{w}_i Y^{\text{obs}}_i - \sum_{i:Z_i = 0}\widehat{w}_i Y^{\text{obs}}_i\} - E_{\mathbb{T}}\{Y(1) - Y(0)\}$ as follows.  
\begin{equation}
    T - \tau = S + R_0 + R_1 + R_2 + \tilde{R}_0 + \tilde{R}_1 + \tilde{R}_2,
\end{equation}
where 
\begin{align*}
    &S = (n^*)^{-1}\sum_{i=1}^{n^*} \{m_1(\bm{X}_i) - m_0(\bm{X}_i)\} + (n^*)^{-1}\sum_{i=1}^{n^*}(D_i Z_i)/\{e(\bm{X}_i) \pi(\bm{X}_i)\}\{Y^{\text{obs}}_i-m_1(\bm{X}_i)\} \\
    &\quad \quad \quad- (n^*)^{-1}\sum_{i=1}^{n^*}(D_i(1-Z_i))/[\{1-e(\bm{X}_i)\}\pi(\bm{X}_i)]\{Y^{\text{obs}}_i-m_0(\bm{X}_i)\} -\tau,\\
    &R_0 = \sum_{i=1}^{n^*}D_iZ_i[\widehat{w}_i - \{n^* e(\bm{X}_i)\pi(\bm{X}_i)\}^{-1}]\{ Y_i(1) - m_1(\bm{X}_i) \},\\ 
&R_1 = \sum_{i=1}^{n^*} (D_iZ_i\widehat{w}_i - (n^*)^{-1})\{m_1(\bm{X}_i) - \bm{\lambda}^{*\top}_{21} \bm{B}(\bm{X}_i)\},\\
&R_2 = \sum_{i=1}^{n^*} (D_iZ_i\widehat{w}_i - (n^*)^{-1}) \{\bm{\lambda}^{*\top}_{21}\bm{B}(\bm{X}_i) \},\\
&\tilde{R}_0 = \sum_{i=1}^{n^*}D_i(1-Z_i)[\widehat{w}_i - \{n^* \{1-e(\bm{X}_i)\}\pi(\bm{X}_i)\}^{-1}]\{ Y_i(0) - m_0(\bm{X}_i) \},\\
&\tilde{R}_1 = \sum_{i=1}^{n^*} \{D_i(1-Z_i)\widehat{w}_i - (n^*)^{-1}\}\{m_0(\bm{X}_i) - \bm{\lambda}^{*\top}_{20} \bm{B}(\bm{X}_i)\},\\
&\tilde{R}_2 = \sum_{i=1}^{n^*} \{D_i(1-Z_i)\widehat{w}_i - (n^*)^{-1}\} \{\bm{\lambda}^{*\top}_{20}\bm{B}(\bm{X}_i) \}.
\end{align*}
By central limit theorem, it follows that,
\begin{equation}
    \sqrt{n^*}S \xrightarrow[n^*\to\infty]{d} \mathcal{N}(0,V),
\end{equation}
where 
\begin{align}
    V &= \Var\Big(m_1(\bm{X}_i) - m_0(\bm{X}_i) + [D_iZ_i\{Y_i(1) - m_1(\bm{X}_i)\}]/\{e(\bm{X}_i)\pi(\bm{X}_i)\} \\
    &\quad \quad \quad - [D_i(1-Z_i)\{Y_i(0) - m_0(\bm{X}_i)\}]/[\{1-e(\bm{X}_i)\}\pi(\bm{X}_i)]   \Big).
\end{align}
$V$ is same as the semiparametric efficiency bound for the target average treatment effect for nested designs \textcolor{black}{(See \citealt{dahabreh2019generalizing}, \citealt{li2021note})}.

For notational convenience, we denote $\tilde{Z}_i = Z_i D_i$. We now consider 
\begin{align}
    \sqrt{n^*}R_0 &= \sqrt{n^*}\Big[\sum_{i=1}^{n^*}\tilde{Z}_i\{\widehat{w}_i - (n^*)^{-1}w^{\text{IP}}_i \}\{ Y_i(1) - m_1(\bm{X}_i) \}\Big] \nonumber \\
    &= \sqrt{n^*}\Big((n^*)^{-1}\sum_{i=1}^{n^*}\tilde{Z}_i\{ Y_i(1) - m_1(\bm{X}_i) \} [n^*\rho'\{\bm{B}(\bm{X}_i)^\top \bm{\lambda}^\dagger\} - w^{\text{IP}}_i ]\Big).
\end{align}
For a function $g_1(\cdot)$, let us define the function
\begin{equation}
    f_0(\tilde{Z}, Y(1), \bm{X}) := \tilde{Z} \{Y(1) - m_1(\bm{X})\} \Big[ n^* \rho'\{g_1(\bm{X})\} - w^{\text{IP}}(\bm{X}) \Big] 
\end{equation}
and the corresponding empirical process $\mathbb{G}_n$ given by,
\begin{equation}
    \mathbb{G}_n(f_0) = (n^*)^{1/2}\Big\{ (n^*)^{-1}\sum_{i=1}^{n^*}f_0(\tilde{Z}_i,Y_i(1), \bm{X}_i) - E\{f_0(\tilde{Z},Y(1), \bm{X})\}  \Big\}. 
\end{equation}
First, $E\{f_0(\tilde{Z},Y(1), \bm{X})\} = 0$ by Assumption 1. Now, consider a class of functions $\mathcal{F}$ defined as
\begin{equation}
    \mathcal{F} = \{f_0: \sup\limits_{\bm{x} \in \mathcal{X}}|g_1(\bm{x}) - g^*_1(\bm{x})| \leq \delta_0\},   
\end{equation}
where $\delta_0 = C[ K^{3/4} \{(\log K)/n^*\}^{1/2} + K^{1 - r_z}]$. From the proof of Theorem 4.2, it follows that $\sup\limits_{\bm{x} \in \mathcal{X}}|\bm{B}(\bm{x})^\top \bm{\lambda}^\dagger - g^*_1(\bm{x})| \leq \delta_0$. Hence, 
\begin{equation}
    \sqrt{n^*}|R_0| \leq \sup\limits_{f_0 \in \mathcal{F}}|\mathbb{G}_n(f_0)| 
\end{equation}
By the Markov inequality, $P(\sup\limits_{f_0 \in \mathcal{F}}|\mathbb{G}_n(f_0)| \geq C) \leq C^{-1} E\{\sup\limits_{f_0 \in \mathcal{F}}|\mathbb{G}_n(f_0)|\}$, for $C>0$. Thus, to show $\sqrt{n^*}|R_0| \xrightarrow[n^* \to \infty]{P} 0$, it is enough to show that $E\{\sup\limits_{f_0 \in \mathcal{F}}|\mathbb{G}_n(f_0)|\} \xrightarrow[n^* \to \infty]{P} 0$. Now, assumptions 2 and \ref{assump4}(a) imply that for $f_0 \in \mathcal{F}$, $|f_0(Z,Y(1),\bm{X})| \leq C'|Y(1) - m_1(\bm{X})|\delta_0$ for a constant $C'>0$. So the function $F_0(Z,Y(1),\bm{X}): = C'|Y(1) - m_1(\bm{X})|\delta_0$ is an envelope of $\mathcal{F}$, with $||F_0||_{P,2} := [E\{F_0\{Z,Y(1),\bm{X}\}^2\}]^{1/2}\leq C\delta_0$ for some $C>0$ by Assumption \ref{assump5}(a). By the maximal inequality (see \citealt{van2000asymptotic} Chapter 19) we have
\begin{equation}
    E\{\sup\limits_{f_0 \in \mathcal{F}}|\mathbb{G}_n(f_0)|\} \lesssim J_{[]}(||F_0||_{P,2}, \mathcal{F}, L_2(P)),
\end{equation}
where $J_{[]}(||F_0||_{P,2}, \mathcal{F}, L_2(P)): = \int_{0}^{||F_0||_{P,2}}[\log N_{[]}\{\epsilon,\mathcal{F},L_2(P)\}]^{1/2}d\epsilon$ is the bracketing integral and $\lesssim$ indicates less than up to a constant. We now use similar steps as in \cite{fan2016improving} and \cite{wang2020minimal} to bound the log of the bracketing number. Define $\mathcal{F}_0 = \{f_0: \sup\limits_{\bm{x}\in \mathcal{X}}|g_1(\bm{x}) - g^*_1(\bm{x})|\leq C\}$ for some constant $C>0$. It follows that, $\log N_{[]}(\epsilon,\mathcal{F},L_2(P)) \lesssim \log N_{[]}(\epsilon,\delta_0\mathcal{F}_0,L_2(P)) =  \log N_{[]}(\epsilon/\delta_0,\mathcal{F}_0,L_2(P)) \lesssim \log N_{[]}(\epsilon/\delta_0,\mathcal{G}_1,L_2(P)) \lesssim (\delta_0 / \epsilon)^{1/k_1}$, where the final inequality holds due to Assumption \ref{assump5}(c). This implies,
\begin{equation}
    J_{[]}(||F_0||_{P,2},\mathcal{F},L_2(P)) \lesssim \int_{0}^{C\delta_0} \Big(\delta_0/\epsilon \Big)^{1/(2k_1)} d_\epsilon \lesssim \delta_0/\{1 - 1/(2k_1)\},
\end{equation}
where in the last step we used $2k_1>1$. The right hand side converges to zero as $n^*$ goes to $\infty$. Thus, $\sqrt{n^*}R_0 \xrightarrow[n^* \to \infty]{P} 0$. Following similar steps, we can show $\sqrt{n^*}\tilde{R}_0 \xrightarrow[n^* \to \infty]{P} 0$.

We will now show that $\sqrt{n^*}R_1 \xrightarrow[n^* \to \infty]{P} 0$ where $R_1 = \sum_{i=1}^{n^*} (\tilde{Z}_i\widehat{w}_i - (n^*)^{-1})\{m_1(\bm{X}_i) - \bm{\lambda}^{*\top}_{21} \bm{B}(\bm{X}_i)\}$. Define the function
\begin{equation}
    f_1(\tilde{Z}, \bm{X}) := [n^*\tilde{Z}\rho'\{g_1(\bm{X})\} - 1]\{m_1(\bm{X}) - \bm{\lambda}^*_{21}\bm{B}(\bm{X})\},
\end{equation}
and the corresponding empirical process $\mathbb{G}_n$ given by,
\begin{equation}
    \mathbb{G}_n(f_1) = (n^*)^{1/2}\Big\{ (n^*)^{-1}\sum_{i=1}^{n^*}f_1(\tilde{Z}_i, \bm{X}_i) - E\{f_1(\tilde{Z},\bm{X})\} \Big\}
\end{equation}
Denote $\Delta(\bm{x}) = m_1(\bm{x}) - \bm{\lambda}^{*\top}_{21}\bm{B}(\bm{x})$. Now, consider a class of functions $\mathcal{F}_1$ defined as,
\begin{equation}
    \mathcal{F}_1 = \{f_1: \sup\limits_{\bm{x} \in \mathcal{X}}|g_1(\bm{x}) - g^*_1(\bm{x})|\leq \delta_1, \sup\limits_{\bm{x} \in \mathcal{X}}|\Delta(\bm{x})|\leq \delta_2  \},
\end{equation} 
where $\delta_1 = C[ K^{3/4} \{(\log K)/n^*\}^{1/2} + K^{1 - r_z}]$, $\delta_2 = CK^{-s_1}$ for some constant $C>0$. As before, using the proof of Theorem 4.2, we get
\begin{equation}
    (n^*)^{1/2}|R_1| \leq \sup\limits_{f_1 \in \mathcal{F}_1}|\mathbb{G}_n(f_1) + \sqrt{n^*}E\{f_1(\tilde{Z},\bm{X})\}|\leq \sup\limits_{f_1 \in \mathcal{F}_1}|\mathbb{G}_n(f_1)| +  (n^*)^{1/2}\sup\limits_{f_1 \in \mathcal{F}_1}|E\{f_1(\tilde{Z},\bm{X})\}|   
\end{equation}
We show that each term on the right hand side is $o_P(1)$. For the first term, by the Markov inequality it suffices to show that $E\{\sup\limits_{f_1 \in \mathcal{F}_1}|\mathbb{G}_n(f_1)|\} \xrightarrow{n^* \to \infty} 0$. Now, assumptions 2, \ref{assump4}(a), and 3 imply that for $f_1 \in \mathcal{F}_1$, $|f_1(Z,\bm{X})| \leq C'\delta_2$ for a constant $C'>0$. So the function $F_1(Z,\bm{X}): = C'\delta_2$ is an envelope of $\mathcal{F}_1$, with $||F_1||_{P,2}\leq C'\delta_2$.
By the maximal inequality,
\begin{equation}
    E\{\sup\limits_{f_1 \in \mathcal{F}_1}|\mathbb{G}_n(f_1)|\} \lesssim J_{[]}(||F_1||_{P,2}, \mathcal{F}_1, L_2(P)),
\end{equation}
where 
\begin{equation}
    J_{[]}(||F_1||_{P,2},\mathcal{F}_1,L_2(P)) \lesssim \int_{0}^{C'\delta_2}\{\log N_{[]}(\epsilon,\mathcal{F}_1,L_2(P))\}^{1/2}d\epsilon. 
\end{equation}
Define $\mathcal{F}_0 := \{f_1: \sup\limits_{\bm{x} \in \mathcal{X}}|g_1(\bm{x}) - g^*_1(\bm{x})|\leq C, \sup\limits_{\bm{x} \in \mathcal{X}}|\Delta(\bm{x})|\leq 1 \}$ for some constant $C>0$,  $\mathcal{H}_{10}:= \{\gamma \in \mathcal{G}_1 + g^*_1: \sup\limits_{\bm{x} \in \mathcal{X}}|\gamma(\bm{x})| \leq C  \}$, $\mathcal{H}_{20}:= \{\Delta \in \mathcal{M}_1 - \bm{\lambda}^{*\top}_{21}\bm{B}(\bm{x}): \sup\limits_{\bm{x} \in \mathcal{X}}|\Delta(\bm{x})| \leq 1  \}$. Using similar steps as in \cite{fan2016improving} and \cite{wang2020minimal} to bound the log of this bracketing number, we get $\log N_{[]}(\epsilon,\mathcal{F}_1,L_2(P)) \lesssim \log N_{[]}(\epsilon/\delta_2,\mathcal{F}_0,L_2(P)) \lesssim \log N_{[]}(\epsilon/\delta_2,\mathcal{H}_{10},L_2(P)) +  \log N_{[]}(\epsilon/\delta_2,\mathcal{H}_{20},L_2(P)) \lesssim  \log N_{[]}(\epsilon/\delta_2,\mathcal{G}_{1},L_2(P)) +  \log N_{[]}(\epsilon/\delta_2,\mathcal{M}_{1},L_2(P)) \leq (\delta_2/\epsilon)^{1/k_1} +(\delta_2/\epsilon)^{1/k_2}$, where the final inequality holds due to assumptions \ref{assump5}(b) and \ref{assump5}(c). Thus, 
\begin{equation}
    J_{[]}(||F_1||_{P,2},\mathcal{F}_1,L_2(P)) \lesssim \delta_2/\{1 - 1/(2k_1)\} + \delta_2 / \{1 - 1/(2 k_2)\}, 
\end{equation}
since $2k_1>1$ and $2k_2>1$. The right hand side converges to zero as $n^*$ goes to infinity. Thus, $\sup\limits_{f_1 \in \mathcal{F}_1}|\mathbb{G}_n(f_1)| = o_P(1)$. 

Now, let $\mathcal{H}_1 = \{g\in \mathcal{G}_1: \sup\limits_{\bm{x} \in \mathcal{X}}|g_1(\bm{x}) - g^*_1(\bm{x})|\leq \delta_1  \}$ and $\mathcal{H}_2 = \{\Delta \in \mathcal{M}_1 - \bm{\lambda}^{*\top}_{21}\bm{B}: \sup\limits_{\bm{x} \in \mathcal{X}}|\Delta(\bm{x})|\leq \delta_2 \}$.  
\begin{align}
   (n^*)^{1/2}\sup\limits_{f_1 \in \mathcal{F}_1}|E\{f_1(\tilde{Z},\bm{X})\}| &= (n^*)^{1/2} \sup\limits_{g_1\in \mathcal{H}_1, \Delta \in \mathcal{H}_2}|E(\{n^*\tilde{Z}\rho'\{g_1(\bm{X})\} - 1\}\Delta(\bm{X}))| \nonumber\\
    &= (n^*)^{1/2}\sup\limits_{g_1\in \mathcal{H}_1, \Delta \in \mathcal{H}_2}\Big|E\Big[\Big( [n^*\rho'\{g_1(\bm{X})\}]/[n^*\rho'\{g^*_1(\bm{X})\}]-1\Big)\Delta(\bm{X})\Big]  \Big| \nonumber\\
    & \lesssim (n^*)^{1/2}\sup\limits_{g_1\in \mathcal{H}_1, \Delta \in \mathcal{H}_2} \Big\{\sup\limits_{\bm{x} \in \mathcal{X}}|g_1(\bm{x}) - g^*_1(\bm{x})| \Big\} \Big\{\sup\limits_{\bm{x} \in \mathcal{X}}|\Delta(\bm{x})|   \Big\} \nonumber\\
  &\lesssim (n^*)^{1/2}\delta_1\delta_2 \nonumber\\
  &\lesssim K^{-s_1 + 3/4} \log K+ (n^*)^{1/2}K^{ - r_1 - s_1 + 1/2}. 
  \label{eq_finalrate}
\end{align}
Here the first inequality holds by applying the mean value theorem and assumptions \ref{assump4}(a), 2, and \ref{assump4}(a). The second inequality holds by definition of $\mathcal{H}_1$ and $\mathcal{H}_2$. Finally, by assumptions 3 and \ref{assump5}(d), the right hand side of Equation \ref{eq_finalrate} goes to zero as $n^*$ goes to infinity. 

Thus, $(n^*)^{1/2}R_1 \xrightarrow[n^* \to \infty]{P} 0$. Following similar steps, we can show $(n^*)^{1/2}\tilde{R}_1 \xrightarrow[n^* \to \infty]{P} 0$. Finally, we consider $R_2 = \sum_{i=1}^{n^*} (\tilde{Z}_i\widehat{w}_i - (n^*)^{-1}) \{\bm{\lambda}^{*\top}_{21}\bm{B}(\bm{X}_i) \}$. We observe that,
\begin{align}
    (n^*)^{1/2}|R_2|&\leq ||\bm{\lambda}^*_{21}||_2  \Big|\Big|\sum_{i:\tilde{Z}_i=1}\widehat{w}_i \bm{B}(\bm{X}_i)- (n^*)^{-1}\sum_{i=1}^{n^*}\bm{B}(\bm{X}_i)\Big|\Big|_2 \nonumber\\
    &\leq ||\bm{\lambda}^*_{21}||_2||\bm{\delta}||_2 = o(1) 
\end{align}
The first inequality is due to Cauchy-Schwarz; the second inequality holds by construction of the weights, and the third equality holds by Assumption \ref{assump3}. This implies, $(n^*)^{1/2}R_2 \xrightarrow[n^* \to \infty]{P} 0$. Following similar steps, we can show $(n^*)^{1/2}\tilde{R}_2 \xrightarrow[n^* \to \infty]{P} 0$. This completes the proof of the theorem.


\subsection{Details for the Simulation Study}
\label{appendix_simulation}

In this section, we include additional results on the performance of the one-step and two-step estimators.
We also include their performance when the estimand is $\mathbb{E}[Y(1) - Y(0) | S = 0]$ (transportability) rather than $\mathbb{E}[Y(1) - Y(0) | S = 1]$ (generalizability).

\textcolor{black}{For the settings in the main text,} Table \ref{tab:TabBias} shows the bias of the H\'{a}jek estimators of the target average treatment effect under each weighting method, based on 800 simulations.
The one-step estimators tend to perform better than the corresponding two-step estimators across the three outcome models, in both the randomized and observational study settings. 
In the randomized study, the one-step estimators reduce the absolute bias by $85\%$ in the misspecified cases and by $79\%$ in the correctly specified cases relative to the corresponding two-step estimators, on average. In the observational study,
the biases are typically higher than their experimental counterparts, particularly for the misspecified cases. Here, the one-step estimators reduce the absolute bias by $10\%$ in the misspecified cases and by $92\%$ in the correctly specified cases relative to the corresponding two-step estimators, on average.
In the correctly specified cases, biases are reduced substantially under the one-step method because by construction, the one-step method balances the right functions of covariates relative to the target.

\begin{table}[H]
\singlespacing
\centering
\caption{\label{tab:TabBias} Bias of the H\'{a}jek estimator of the target average treatment effect using different weighting methods in both the randomized and observational study settings.}
\begin{tabular}{lrrrrrrr}
                  & \multicolumn{3}{c}{Randomized Study Setting} & \multicolumn{1}{l}{} & \multicolumn{3}{c}{Observational Study Setting} \\ \cline{2-4} \cline{6-8} 
                  & Outcome       & Outcome       & Outcome      &                      & Outcome        & Outcome        & Outcome       \\
Weighting Method  & Model 1       & Model 2       & Model 3      &                      & Model 1        & Model 2        & Model 3       \\ \hline
Two-Step Method 1 & 0.67          & 3.30          & 5.02         &                      & -9.26          & -6.45          & -4.77         \\
One-Step Method 1 & -0.02         & 0.53          & 2.04         &                      & -8.62          & -8.00          & -6.46         \\
Two-Step Method 2 & 0.79          & 3.59          & 3.15         &                      & -32.12         & -27.84         & -32.71        \\
One-Step Method 2 & -0.03         & 0.50          & -0.43        &                      & -17.16         & -16.21         & -23.97        \\
Two-Step Method 3 & 0.18          & 0.20          & 0.14         &                      & 0.63           & 0.91           & 1.05          \\
One-Step Method 3 & 0.01          & -0.03         & -0.06        &                      & -0.10          & -0.04          & 0.03   \\      
\hline
\end{tabular}
\end{table}

\color{black}
For the transportability design, we slightly modify the setup in Section 5.
There are $n_{\text{study}} = 500$ units in the study sample and $n_{\text{target}} = 5000$ units in the target sample.
Let $S = 1$ indicate a unit is in the study sample and $S = 0$ indicate a unit is in the target sample.
For units with $S = 1$, the four independent latent covariates are $U_1$, $U_2$, $U_3$, $U_4 \sim \mathcal{N}(0, 1)$.
For units with $S = 0$, they are $U_1 \sim \mathcal{N}(1.2, 1)$, $U_2 \sim \mathcal{N}(-0.4, 1)$, $U_3 \sim \mathcal{N}(0.3, 1)$, and $U_4 \sim \mathcal{N}(0.1, 1)$.
This creates similar correlations betwen the latent covariates and $S$ as in the setting detailed in Section 5.
The remaining setup is the same as that in Section 5.

\begin{centering}
\begin{figure}[H]
\centering
\caption{\textcolor{black}{Effective sample sizes and maximum normalized weights across weighting methods in both the randomized and observational study settings (transportability setting).}}\label{fig:ess2a}
\includegraphics[scale=0.9]{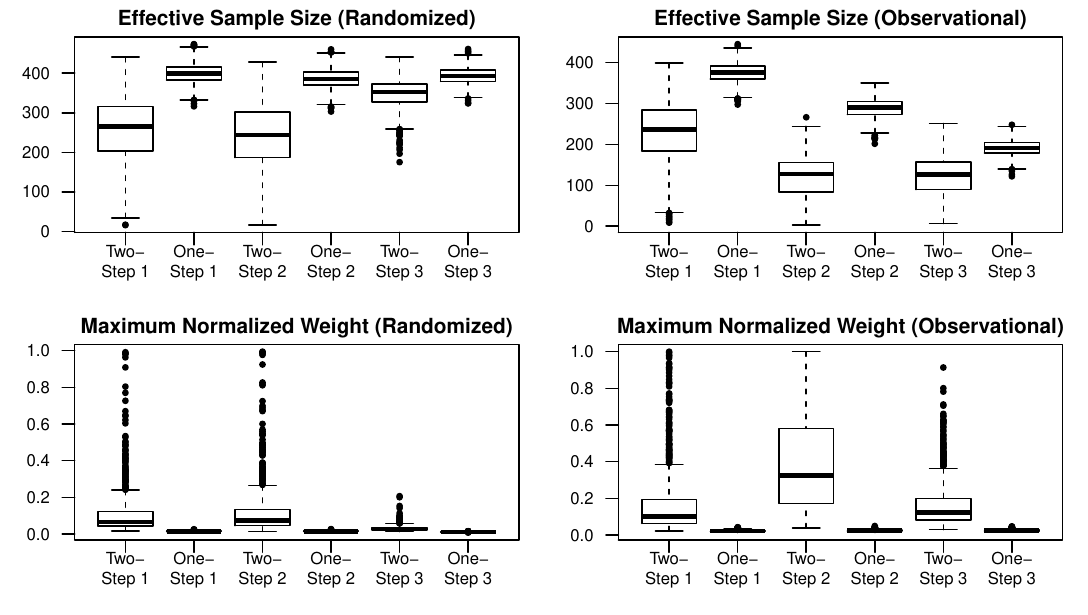}
\end{figure}
\end{centering}

Table \ref{tab:tab2a} shows the root-mean-squared errors and Table \ref{tab:Tab3a} shows the mean bias of the H\'{a}jek estimators of the target average treatment effect.
Figure \ref{fig:ess2a} shows how the effective sample sizes and maximum normalized weights vary across simulations.
Overall, the pattern of results is similar as in Section 5, albeit with a more dramatic improvement in performance by the one-step weights, likely due to the more limited covariate overlap.
A few exceptions occur for the observational study, where the two-step weights produce less biased estimates than the one-step weights for outcome model 1 (i.e., no treatment effect heterogeneity), though the one-step weights still show improved root-mean-squared error.

\begin{table}[H]
\color{black}
\singlespacing
\centering
\caption{\textcolor{black}{\label{tab:tab2a} Root-mean-squared error of the H\'{a}jek estimator of the target average treatment effect using different weighting methods in both the randomized and observational study settings (transportability setting).}}
\begin{tabular}{lrrrrrrr}
                 & \multicolumn{3}{c}{Randomized Study Setting}                                            & \multicolumn{1}{r}{} & \multicolumn{3}{c}{Observational Study Setting}                                         \\ \cline{2-4}
                 \cline{6-8} 
                 & \multicolumn{1}{r}{Outcome} & \multicolumn{1}{r}{Outcome} & \multicolumn{1}{r}{Outcome} & \multicolumn{1}{r}{} & \multicolumn{1}{r}{Outcome} & \multicolumn{1}{r}{Outcome} & \multicolumn{1}{r}{Outcome} \\
Weighting Method & \multicolumn{1}{r}{Model 1} & \multicolumn{1}{r}{Model 2} & \multicolumn{1}{r}{Model 3} & \multicolumn{1}{r}{} & \multicolumn{1}{r}{Model 1} & \multicolumn{1}{r}{Model 2} & \multicolumn{1}{r}{Model 3} \\ \hline
Two-Step 1       & 30.35                       & 45.61                       & 48.11                       &                      & 29.56                       & 45.78                       & 47.33                       \\
One-Step 1       & 2.32                        & 2.52                        & 3.24                        &                      & 5.64                        & 5.15                        & 5.59                        \\
Two-Step 2       & 31.88                       & 46.71                       & 49.74                       &                      & 31.90                       & 46.02                       & 48.44                       \\
One-Step 2       & 1.56                        & 1.81                        & 2.42                        &                      & 14.05                       & 12.98                       & 20.97                       \\
Two-Step 3       & 8.81                        & 19.15                       & 18.57                       &                      & 12.81                       & 22.27                       & 23.20                       \\
One-Step 3       & 0.52                        & 0.81                        & 1.22                        &                      & 0.72                        & 0.95                        & 1.33                        \\ \hline
\end{tabular}
\end{table}

\begin{table}[H]
\color{black}
\singlespacing
\centering
\caption{\label{tab:Tab3a}\textcolor{black}{Bias of the H\'{a}jek estimator of the target average treatment effect using different weighting methods in both the randomized and observational study settings (transportability setting).}}
\begin{tabular}{lrrrrrrr}
                 & \multicolumn{3}{c}{Randomized Study Setting}                                            & \multicolumn{1}{r}{} & \multicolumn{3}{c}{Observational Study Setting}                                         \\ \cline{2-4} \cline{6-8} 
                 & \multicolumn{1}{r}{Outcome} & \multicolumn{1}{r}{Outcome} & \multicolumn{1}{r}{Outcome} & \multicolumn{1}{r}{} & \multicolumn{1}{r}{Outcome} & \multicolumn{1}{r}{Outcome} & \multicolumn{1}{r}{Outcome} \\
Weighting Method & \multicolumn{1}{r}{Model 1} & \multicolumn{1}{r}{Model 2} & \multicolumn{1}{r}{Model 3} & \multicolumn{1}{r}{} & \multicolumn{1}{r}{Model 1} & \multicolumn{1}{r}{Model 2} & \multicolumn{1}{r}{Model 3} \\ \hline
Two-Step 1       & -0.29                       & 27.29                       & 55.09                       &                      & -0.37                       & 28.08                       & 55.13                       \\
One-Step 1       & -0.06                       & -0.07                       & -0.07                       &                      & -5.12                       & -4.47                       & 3.21                        \\
Two-Step 2       & -0.38                       & 26.87                       & 54.85                       &                      & -3.39                       & 24.10                       & 55.07                       \\
One-Step 2       & -0.02                       & 0.00                        & -0.04                       &                      & -13.95                      & -12.83                      & 7.69                        \\
Two-Step 3       & -0.27                       & 16.04                       & 32.80                       &                      & 0.03                        & 15.92                       & 31.17                       \\
One-Step 3       & -0.02                       & -0.02                       & -0.09                       &                      & 0.21                        & 0.29                        & -0.06                       \\ \hline
\end{tabular}
\end{table}

\subsection{Simulation study with heteroscedastic outcomes}
\label{sec_simu_hetro}

In this section, we compare the one-step and two-step estimators in settings where the potential outcomes are heteroscedastic. 
We use the same setting as in Section \ref{sec_simulation} with four independent unobserved covariates distributed as $U_1, U_2, U_3, U_4 \sim \mathcal{N}(0, 1)$, and four observed covariates generated as $X_1 = \exp(U_1/2)$, $X_2 = U_2/\{1 + \exp(U_1)\} + 10$, $X_3 = (U_1U_3/25 + 0.6)^3$, and $X_4 = (U_2 + U_4 + 20)^{2}$.
$D$ is the binary indicator for selection into the study, and $Z$ is the binary treatment indicator.
The true model for the probability of selection into the study is $\mathrm{pr}(D = 1 | \bm{U}) = \text{expit}(-U_1 + 0.5 U_2 - 0.25 U_3 - 0.1 U_4)$ so that, marginally, $\mathrm{pr}(D = 1)  = 0.5$. 
The total cohort size is 1000.
For the randomized study setting, $\mathrm{pr}(Z = 1|\bm{U}) = 0.5$, and for the observational setting, $\mathrm{pr}(Z=1|\bm{U}) = \text{expit}(U_1 + 2U_2 - 2U_3 - U_4)$.

We consider three different models for $Y(0)$ and $Y(1)$. Under Model-$j$ ($j \in \{1,2,3\}$), $Y(0) = 210 + 27.4 U_1 + 13.7 U_2 + 13.7 U_3 + 13.7 U_4 + h_j(\bm{U})\epsilon_0$, where $\epsilon_0 \sim \mathcal{N}(0, 5^2)$. This model allows for heteroscedasticity, since the conditional variance varies as a function of the covariates, i.e., $\Var_{\mathbb{T}}(Y(0)|\bm{U} = \bm{u}) = 25 h^2_j(\bm{u})$. We set $h_1(\bm{u}) = 2u_1$, $h_2(\bm{u}) = 2(u_1 + u_2)$, and $h_3(\bm{u}) = 2(u_1 + u_2 + u_3 + u_4)$. Similarly, there are three models for $Y(1)$: Model 1 is given by $Y(1) = 210 + 27.4 U_1 + 13.7 U_2 + 13.7 U_3 + 13.7 U_4 + h_1(\bm{U})\epsilon_1$; Model 2 by $Y(1) = 210 + 41.1 U_1 + 13.7 U_2 + 13.7 U_3 + 13.7 U_4 + h_2(\bm{U})\epsilon_1$; and Model 3 by $Y(1) = 210 + 41.1 U_1 + 27.4 U_2 + 27.4 U_3 + 13.7 U_4 + h_3(\bm{U})\epsilon_1$; where $\epsilon_1 \sim \mathcal{N}(0,5^2)$ and $h_j(\cdot)$s are the same as those for the models of $Y(0)$.

We compare three versions of one-step weighting to three versions of two-step weighting as specified in Section \ref{sec_simulation}. Table \ref{tab:hetero} shows the root-mean-squared errors of the H\'{a}jek estimators, based on 800 simulations. We observe that the one-step weights outperform the two-step weights across the three outcome models in both the randomized and observational study settings.

\begin{table}[H]
\singlespacing
\centering
\caption{\label{tab:hetero} Root-mean-squared error of the H\'{a}jek estimator of the target average treatment effect using different weighting methods in both the randomized and observational study settings under heteroscedastic outcomes.}
\begin{tabular}{lrrrrrrr}
                 & \multicolumn{3}{c}{Randomized Study Setting}                                            & \multicolumn{1}{r}{} & \multicolumn{3}{c}{Observational Study Setting}                                         \\ \cline{2-4}
                 \cline{6-8} 
                 & \multicolumn{1}{r}{Outcome} & \multicolumn{1}{r}{Outcome} & \multicolumn{1}{r}{Outcome} & \multicolumn{1}{r}{} & \multicolumn{1}{r}{Outcome} & \multicolumn{1}{r}{Outcome} & \multicolumn{1}{r}{Outcome} \\
Weighting Method & \multicolumn{1}{r}{Model 1} & \multicolumn{1}{r}{Model 2} & \multicolumn{1}{r}{Model 3} & \multicolumn{1}{r}{} & \multicolumn{1}{r}{Model 1} & \multicolumn{1}{r}{Model 2} & \multicolumn{1}{r}{Model 3} \\ \hline
Two-Step 1 & 19.23  & 23.21  & 26.33 & & 23.83 & 24.96 & 25.72  \\ 
One-Step 1 & 3.03  & 3.57  & 4.75 & & 9.23 & 8.69 & 7.93  \\
Two-Step 2 &  20.83 & 25.25 & 28.09 & & 44.45 & 42.65 & 48.17\\
One-Step 2 &  2.42  & 3.02  & 3.84 & & 17.34 & 16.41 & 24.41\\
Two-Step 3 &  5.07  & 5.73  & 6.58 & & 9.68 & 12.64  & 13.92\\
One-Step 3 & 0.89 & 1.52 & 2.03 & & 1.22  & 1.73  & 2.99 \\
\bottomrule
\end{tabular}
\end{table}

\color{black}
\subsection{Additional Case Study Results}
In this section, we present results from the case study in Section 5, albeit with the weights implemented via the two-step method.
To calculate the two-step weights, we fit logistic regression models for treatment and study selection, and we trim each set of weights at their $90$th percentiles.

\begin{centering}
\begin{figure}[H]
\color{black}
\singlespacing
\caption{Distributions of target absolute standardized mean differences and effective sample sizes for three target populations (two-step).}\label{fig:fig1IOW}
\includegraphics[scale=0.7]{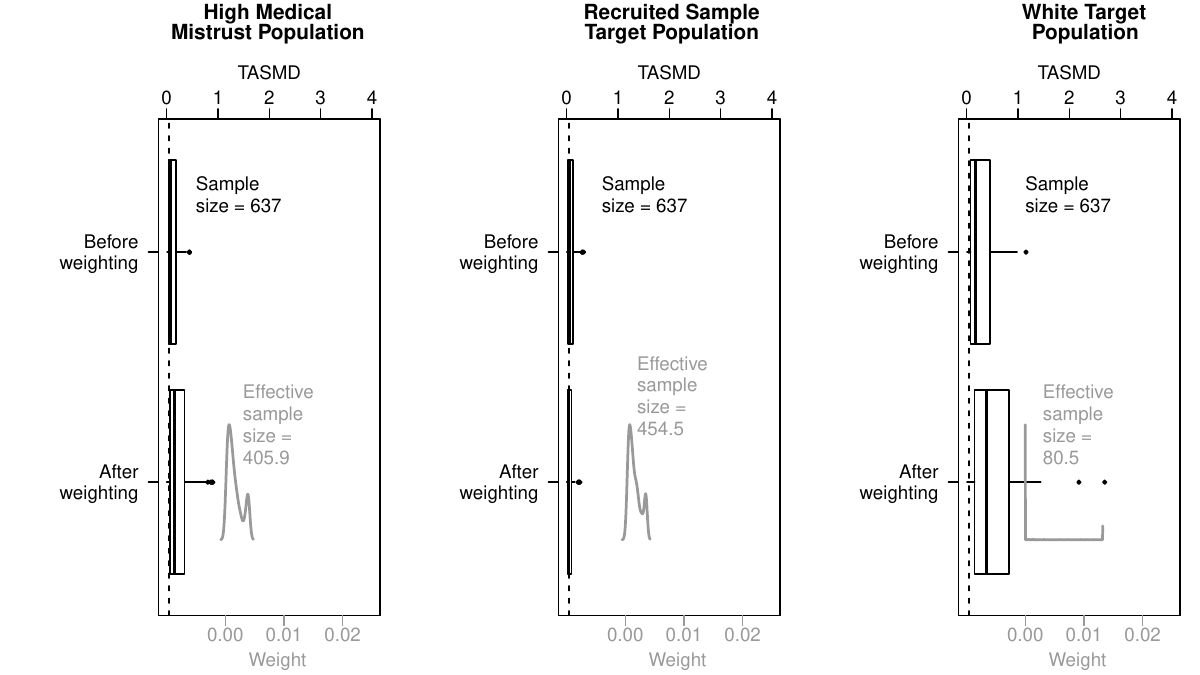}
\begin{flushleft}
\scriptsize{TASMD = target absolute standardized mean difference. The black vertical dashed line in each plot marks a TASMD of 0.05, signifying the heuristic that a TASMD $<$ 0.05 indicates good balance. SD = standard deviation.}
\end{flushleft}
\end{figure}
\end{centering}

Figure \ref{fig:fig1IOW} summarizes the performance of the two-step weighting method for achieving balance relative to the various target covariate profiles.
The figure also summarizes the dispersion of the weights via density plots and effective sample sizes.
Compared to the weights in Figure \ref{fig:fig1}, each set of two-step weights is less evenly dispersed and has higher variance and lower effective sample size, reflecting the one-step method's explicit optimization for these criteria.
Covariate balance is also worse than in the one-step approach, again because the one-step weights explicitly target covariate balance.
This pattern becomes more stark as the profile becomes more difficult to target (i.e., as there is less overlap between the target and study populations).

\begin{figure}[H]
\color{black}
\singlespacing
\centering
\caption{\textcolor{black}{Estimates of the target average treatment effect for various outcome variables and target populations.}}\label{fig:fig2IOW}
\includegraphics[scale=0.75]{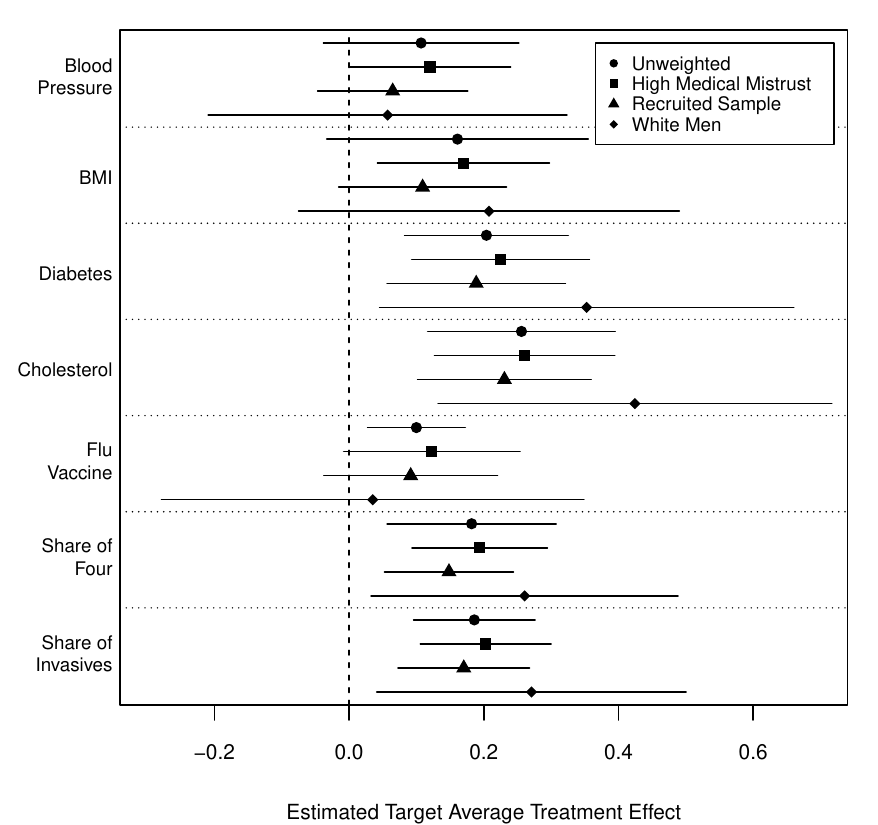}
\end{figure}

Figure \ref{fig:fig2IOW} presents the H\'{a}jek estimates of the target average treatment effect for each outcome along with bootstrapped confidence intervals.
The results are similar as those due to the one-step weights in Figure \ref{fig:fig2}, however, due to the higher variability of the two-step weights, the confidence intervals are much wider.

\color{black}

\subsection{\texttt{R} Code}
In this section, we provide instructions on how to implement the one-step weights using \texttt{sbw} package for \texttt{R}. First, however, we recommend installing \texttt{gurobi}, an optimizer which increases the performance of \texttt{sbw}. Instructions on installation can be found on \url{https://www.gurobi.com}.

Next, install the \texttt{sbw} package in \texttt{R} via the code \texttt{install.packages("sbw")}. Now, in this section, we provide example code that would be used to weight the data set from the applied study corresponding to the randomized participants. We weight these data toward a covariate profile constructed from the covariate means of the entire recruited sample.

First, read in the trial data (\texttt{oakland\_analysis\_final.dta}) and the recruited sample data\\ (\texttt{oakland\_analysis\_selection.dta}) and list the covariates to balance. 

\singlespacing
\texttt{> library(sbw)\\
> oakland.final <- read\_dta("oakland\_analysis\_final.dta")\\
> oakland.selection <- read\_dta("oakland\_analysis\_selection.dta")\\
> T1.vars <- c("good\_sa\_health",\\
> \hspace{75pt} "any\_health\_prob",\\
> \hspace{75pt} "ER\_2years",\\
> \hspace{75pt} "nights\_hosp\_2years",\\
> \hspace{75pt} "hosp\_visits\_2years",\\
> \hspace{75pt} "med\_mistrust",\\
> \hspace{75pt} "has\_PCP",\\
> \hspace{75pt} "uninsured",\\
> \hspace{75pt} "age",\\
> \hspace{75pt} "married",\\
> \hspace{75pt} "unemployed",\\
> \hspace{75pt} "HSless",\\
> \hspace{75pt} "low\_income",\\
> \hspace{75pt} "benefits")
}

\doublespacing
For the sake of this analyses, we have imputed missing values with the mean for continuous covariates and created an additional missing category for categorical covariates. For continuous imputed covariates, we also add a dummy variable indicating missingness. These additional variables are added to the list to balance. For the sake of space, we omit including the code that performs these imputations. We assume that the final list of covariates is included in the list \texttt{T1.vars.imp}. We assume the data sets \texttt{oakland.selection} and \texttt{oakland.final} have been updated accordingly.

Next, we define the balance tolerances. We define each covariate's tolerance as 0.1 times the covariate standard deviation in the recruited sample.

\singlespacing
\texttt{> sd\_targets <- apply(as.matrix(oakland.selection[T1.vars.imp]), 2, sd)\\
> tols <- 0.1 \* sd\_targets}

\doublespacing
Next we subset the trial data by treatment group (i.e., values of \texttt{black\_dr}) --- as we want to weight each treatment group toward the target profile. Then we define the various inputs to the \texttt{sbw} function. For this first implementation, we manually set the tolerances, restrict the weights to be positive (\texttt{wei\_pos = TRUE}), and restrict the weights to sum to one (\texttt{wei\_sum = TRUE}).

\singlespacing
\texttt{> t\_ind <- "black\_dr"\\
> dat.1 <- subset(dat, dat[t\_ind][,1] == 1)\\
> dat.0 <- subset(dat, dat[t\_ind][,1] == 0)\\
> bal\_cov <- T1.vars.imp\\
> bal\_alg <- FALSE\\
> bal\_tol <- tols\\
> bal\_std <- "manual"\\
> bal <- list(bal\_cov = bal\_cov, bal\_alg = bal\_alg, bal\_tol = bal\_tol, bal\_std = bal\_std)\\
> wei <- list(wei\_sum = TRUE, wei\_pos = TRUE)\\
> par\_tar <- colMeans(oakland.selection[T1.vars.imp])}

\doublespacing
Next, we balance each treatment group and combine the weighted data into a single data set with both treatment groups.

\singlespacing
\texttt{> sbw.results.1 <- sbw(dat = dat.1, bal = bal, par = list(par\_est = "aux", par\_tar = par\_tar), sol = list(sol\_nam = "gurobi"),wei = wei)\\
> sbw.results.0 <- sbw.results.1 <- sbw(dat = dat.0, bal = bal, par = list(par\_est = "aux", par\_tar = par\_tar), sol = list(sol\_nam = "gurobi"),wei = wei)\\
> weighted.df.1 <- sbw.results.1\$dat\_weights\\
> weighted.df.0 <- sbw.results.0\$dat\_weights\\
> weighted.df <- rbind(weighted.df.1, weighted.df.0)}

\doublespacing
Using the weighted data, we can compute the TATE directly via weighted means. For the sake of demonstration, we evaluate the TATE for the outcome that measures whether the participant elected to receive a flu shot after their doctor's visit.

\singlespacing
\texttt{> mean.1 <- weighted.mean(weighted.df.1["post\_flu"][,1], weighted.df.1["sbw\_weights"][,1])\\
> mean.0 <- weighted.mean(weighted.df.0["post\_flu"][,1], weighted.df.0["sbw\_weights"][,1])}

\doublespace
We omit the code to compute standard errors and confidence intervals via bootstrapping. In addition to supplying the tolerances manually, one could also implement the algorithm that selects the tolerances from a grid of options in a data-adaptive way. Code to implement this method appears below.

\singlespacing
\texttt{> t\_ind <- "black\_dr"\\
> dat.1 <- subset(dat, dat[t\_ind][,1] == 1)\\
> dat.0 <- subset(dat, dat[t\_ind][,1] == 0)\\
> bal\_cov <- T1.vars.imp\\
> bal\_alg <- TRUE\\
> bal\_std <- "manual"\\
> bal\_gri <- c(0.0001, 0.001, 0.002, 0.005, 0.01, 0.02, 0.05, 0.1) \\
> bal <- list(bal\_cov = bal\_cov, bal\_alg = bal\_alg, bal\_tol = NULL, bal\_std = bal\_std, bal\_gri = bal\_gri)\\
> wei <- list(wei\_sum = TRUE, wei\_pos = TRUE)\\
> par\_tar <- colMeans(oakland.selection[T1.vars.imp])\\
> sbw.results.1 <- sbw(dat = dat.1, bal = bal, par = list(par\_est = "aux", par\_tar = par\_tar), sol = list(sol\_nam = "gurobi"),wei = wei)\\
> sbw.results.0 <- sbw.results.1 <- sbw(dat = dat.0, bal = bal, par = list(par\_est = "aux", par\_tar = par\_tar), sol = list(sol\_nam = "gurobi"),wei = wei)\\
> weighted.df.1 <- sbw.results.1\$dat\_weights\\
> weighted.df.0 <- sbw.results.0\$dat\_weights\\
> weighted.df <- rbind(weighted.df.1, weighted.df.0)}




%% file: onestep_arxiv_ec.bbl
\begin{thebibliography}{38}
\newcommand{\enquote}[1]{``#1''}
\expandafter\ifx\csname natexlab\endcsname\relax\def\natexlab#1{#1}\fi

\bibitem[{Alsan et~al.(2019)Alsan, Garrick, and
  Graziani}]{AlsanMarcella2019DDMf}
Alsan, M., Garrick, O., and Graziani, G. (2019), \enquote{Does Diversity Matter
  for Health? Experimental Evidence from Oakland,} \textit{The American
  Economic Review}, 109, 4071--4111.

\bibitem[{Andrews(1991)}]{andrews1991asymptotic}
Andrews, D.~W. (1991), \enquote{Asymptotic normality of series estimators for
  nonparametric and semiparametric regression models,} \textit{Econometrica:
  Journal of the Econometric Society}, 307--345.

\bibitem[{Austin(2016)}]{austin2016variance}
Austin, P.~C. (2016), \enquote{Variance estimation when using inverse
  probability of treatment weighting (IPTW) with survival analysis,}
  \textit{Statistics in Medicine}, 35, 5642--5655.

\bibitem[{Ben-Michael et~al.(2021)Ben-Michael, Feller, Hirshberg, and
  Zubizarreta}]{ben2021balancing}
Ben-Michael, E., Feller, A., Hirshberg, D.~A., and Zubizarreta, J.~R. (2021),
  \enquote{The Balancing Act in Causal Inference,} \textit{arXiv preprint
  arXiv:2110.14831}.

\bibitem[{Boyd and Vandenberghe(2004)}]{boyd2004convex}
Boyd, S. and Vandenberghe, L. (2004), \textit{Convex optimization}, Cambridge
  university press.

\bibitem[{Bruns-Smith et~al.(2023)Bruns-Smith, Dukes, Feller, and
  Ogburn}]{bruns2023augmented}
Bruns-Smith, D., Dukes, O., Feller, A., and Ogburn, E.~L. (2023),
  \enquote{Augmented balancing weights as linear regression,} \textit{arXiv
  preprint arXiv:2304.14545}.

\bibitem[{Chattopadhyay et~al.(2020)Chattopadhyay, Hase, and
  Zubizarreta}]{chattopadhyay2020balancing}
Chattopadhyay, A., Hase, C.~H., and Zubizarreta, J.~R. (2020),
  \enquote{Balancing vs modeling approaches to weighting in practice,}
  \textit{Statistics in Medicine}, 39, 3227--3254.

\bibitem[{Chattopadhyay and Zubizarreta(2023)}]{chattopadhyay2023implied}
Chattopadhyay, A. and Zubizarreta, J.~R. (2023), \enquote{On the implied
  weights of linear regression for causal inference,} \textit{Biometrika}, in
  press.

\bibitem[{Dahabreh et~al.(2020)Dahabreh, Robertson, Steingrimsson, Stuart, and
  Hernan}]{dahabreh2020extending}
Dahabreh, I.~J., Robertson, S.~E., Steingrimsson, J.~A., Stuart, E.~A., and
  Hernan, M.~A. (2020), \enquote{Extending inferences from a randomized trial
  to a new target population,} \textit{Statistics in medicine}, 39, 1999--2014.

\bibitem[{Dahabreh et~al.(2019)Dahabreh, Robertson, Tchetgen, Stuart, and
  Hern{\'a}n}]{dahabreh2019generalizing}
Dahabreh, I.~J., Robertson, S.~E., Tchetgen, E.~J., Stuart, E.~A., and
  Hern{\'a}n, M.~A. (2019), \enquote{Generalizing causal inferences from
  individuals in randomized trials to all trial-eligible individuals,}
  \textit{Biometrics}, 75, 685--694.

\bibitem[{Degtiar and Rose(2023)}]{degtiar2023review}
Degtiar, I. and Rose, S. (2023), \enquote{A Review of Generalizability and
  Transportability,} \textit{Annual Review of Statistics and Its Application},
  10, 501--524.

\bibitem[{Fan et~al.(2016)Fan, Imai, Liu, Ning, and Yang}]{fan2016improving}
Fan, J., Imai, K., Liu, H., Ning, Y., and Yang, X. (2016), \enquote{Improving
  covariate balancing propensity score: A doubly robust and efficient
  approach,} Tech. rep., Princeton University.

\bibitem[{Hainmueller(2012)}]{hainmueller2012balancing}
Hainmueller, J. (2012), \enquote{Entropy balancing for causal effects: a
  multivariate reweighting method to produce balanced samples in observational
  studies,} \textit{Political Analysis}, 20, 25--46.

\bibitem[{Hirshberg and Wager(2021)}]{hirshberg2021augmented}
Hirshberg, D.~A. and Wager, S. (2021), \enquote{Augmented minimax linear
  estimation,} \textit{The Annals of Statistics}, 49, 3206--3227.

\bibitem[{Imbens and Rubin(2015)}]{imbens2015causal}
Imbens, G.~W. and Rubin, D.~B. (2015), \textit{Causal inference in statistics,
  social, and biomedical sciences}, Cambridge University Press.

\bibitem[{Josey et~al.(2021)Josey, Berkowitz, Ghosh, and
  Raghavan}]{josey2021transporting}
Josey, K.~P., Berkowitz, S.~A., Ghosh, D., and Raghavan, S. (2021),
  \enquote{Transporting experimental results with entropy balancing,}
  \textit{Statistics in Medicine}, 40, 4310--4326.

\bibitem[{Kang and Schafer(2007)}]{kang2007demystifying}
Kang, J. D.~Y. and Schafer, J.~L. (2007), \enquote{{D}emystifying double
  robustness: a comparison of alternative strategies for estimating a
  population mean from incomplete data (with discussion),} \textit{Statistical
  Science}, 22, 523--539.

\bibitem[{Kern et~al.(2016)Kern, Stuart, Hill, and Green}]{kern2016assessing}
Kern, H.~L., Stuart, E.~A., Hill, J., and Green, D.~P. (2016),
  \enquote{Assessing methods for generalizing experimental impact estimates to
  target populations,} \textit{Journal of Research on Educational
  Effectiveness}, 9, 103--127.

\bibitem[{Lee et~al.(2021)Lee, Yang, Dong, Wang, Zeng, and
  Cai}]{lee2021improving}
Lee, D., Yang, S., Dong, L., Wang, X., Zeng, D., and Cai, J. (2021),
  \enquote{Improving trial generalizability using observational studies,}
  \textit{Biometrics}.

\bibitem[{Li et~al.(2021)Li, Hong, and Stuart}]{li2021note}
Li, F., Hong, H., and Stuart, E.~A. (2021), \enquote{A note on semiparametric
  efficient generalization of causal effects from randomized trials to target
  populations,} \textit{Communications in Statistics-Theory and Methods},
  1--32.

\bibitem[{Li et~al.(2018)Li, Morgan, and Zaslavsky}]{li2018balancing}
Li, F., Morgan, K.~L., and Zaslavsky, A.~M. (2018), \enquote{Balancing
  covariates via propensity score weighting,} \textit{Journal of the American
  Statistical Association}, 113, 390--400.

\bibitem[{Lipsitch et~al.(2010)Lipsitch, Tchetgen~Tchetgen, and
  Cohen}]{LipsitchMarc2010NCAT}
Lipsitch, M., Tchetgen~Tchetgen, E., and Cohen, T. (2010), \enquote{Negative
  Controls: A Tool for Detecting Confounding and Bias in Observational
  Studies,} \textit{Epidemiology (Cambridge, Mass.)}, 21, 383--388.

\bibitem[{Lu et~al.(2021)Lu, Ben-Michael, Feller, and Miratrix}]{lu2021you}
Lu, B., Ben-Michael, E., Feller, A., and Miratrix, L. (2021), \enquote{Is it
  who you are or where you are? Accounting for compositional differences in
  cross-site treatment variation,} \textit{arXiv preprint arXiv:2103.14765}.

\bibitem[{Newey(1997)}]{newey1997convergence}
Newey, W.~K. (1997), \enquote{Convergence rates and asymptotic normality for
  series estimators,} \textit{Journal of Econometrics}, 79, 147--168.

\bibitem[{Neyman(1923)}]{neyman1923application}
Neyman, J. (1923), \enquote{On the application of probability theory to
  agricultural experiments,} \textit{Statistical Science}, 5, 463--480.

\bibitem[{Robins et~al.(2007)Robins, Sued, Lei-Gomez, and
  Rotnitzky}]{robins2007comment}
Robins, J., Sued, M., Lei-Gomez, Q., and Rotnitzky, A. (2007),
  \enquote{Comment: performance of double-robust estimators when" inverse
  probability" weights are highly variable,} \textit{Statistical Science}, 22,
  544--559.

\bibitem[{Robins and Rotnitzky(1995)}]{robins1995semiparametric}
Robins, J.~M. and Rotnitzky, A. (1995), \enquote{Semiparametric efficiency in
  multivariate regression models with missing data,} \textit{Journal of the
  American Statistical Association}, 90, 122--129.

\bibitem[{Rosenbaum and Rubin(1983)}]{rosenbaum1983central}
Rosenbaum, P.~R. and Rubin, D.~B. (1983), \enquote{The central role of the
  propensity score in observational studies for causal effects,}
  \textit{Biometrika}, 70, 41--55.

\bibitem[{Rubin(1974)}]{rubin1974estimating}
Rubin, D.~B. (1974), \enquote{Estimating causal effects of treatments in
  randomized and nonrandomized studies.} \textit{Journal of Educational
  Psychology}, 66, 688.

\bibitem[{Stuart et~al.(2011)Stuart, Cole, Bradshaw, and Leaf}]{stuart2011use}
Stuart, E.~A., Cole, S.~R., Bradshaw, C.~P., and Leaf, P.~J. (2011),
  \enquote{The use of propensity scores to assess the generalizability of
  results from randomized trials,} \textit{Journal of the Royal Statistical
  Society: Series A}, 174, 369--386.

\bibitem[{Tipton and Olsen(2018)}]{tipton2018review}
Tipton, E. and Olsen, R.~B. (2018), \enquote{A review of statistical methods
  for generalizing from evaluations of educational interventions,}
  \textit{Educational Researcher}, 47, 516--524.

\bibitem[{Tropp et~al.(2015)}]{tropp2015introduction}
Tropp, J.~A. et~al. (2015), \enquote{An introduction to matrix concentration
  inequalities,} \textit{Foundations and Trends{\textregistered} in Machine
  Learning}, 8, 1--230.

\bibitem[{Van~der Vaart(2000)}]{van2000asymptotic}
Van~der Vaart, A.~W. (2000), \textit{Asymptotic statistics}, vol.~3, Cambridge
  university press.

\bibitem[{Wang and Zubizarreta(2020)}]{wang2020minimal}
Wang, Y. and Zubizarreta, J.~R. (2020), \enquote{Minimal dispersion
  approximately balancing weights: asymptotic properties and practical
  considerations,} \textit{Biometrika}, 107, 93--105.

\bibitem[{Westreich et~al.(2017)Westreich, Edwards, Lesko, Stuart, and
  Cole}]{westreich2017transportability}
Westreich, D., Edwards, J.~K., Lesko, C.~R., Stuart, E., and Cole, S.~R.
  (2017), \enquote{Transportability of trial results using inverse odds of
  sampling weights,} \textit{American Journal of Epidemiology}, 186,
  1010--1014.

\bibitem[{Zhao and Percival(2016)}]{zhao2016entropy}
Zhao, Q. and Percival, D. (2016), \enquote{Entropy balancing is doubly robust,}
  .

\bibitem[{Zhao et~al.(2019)Zhao, Small, and Bhattacharya}]{zhao2019sensitivity}
Zhao, Q., Small, D.~S., and Bhattacharya, B.~B. (2019), \enquote{Sensitivity
  analysis for inverse probability weighting estimators via the percentile
  bootstrap,} \textit{Journal of the Royal Statistical Society: Series B
  (Statistical Methodology)}, 81, 735--761.

\bibitem[{Zubizarreta(2015)}]{zubizarreta2015stable}
Zubizarreta, J.~R. (2015), \enquote{Stable weights that balance covariates for
  estimation with incomplete outcome data,} \textit{Journal of the American
  Statistical Association}, 110, 910--922.

\end{thebibliography}
